\documentclass[aps,superscriptaddress,twocolumn]{revtex4-1}

\usepackage[english]{babel}
\usepackage{upgreek}
\usepackage{amsmath,amsfonts,amsthm,amsbsy,empheq}
\usepackage{mathtools}
\usepackage[caption=false]{subfig}
\usepackage{graphicx,bm}
\usepackage{physics}
\usepackage{dsfont}
\usepackage{bibentry}
\usepackage{xcolor}
\usepackage{breqn}
\usepackage{ulem}
\definecolor{dark-green}{RGB}{0, 128, 0}

\DeclareTextCompositeCommand{\k}{LY1}{e}
{\oalign{e\crcr\noalign{\kern-.27ex}\hidewidth\char7\hidewidth}}

\usepackage[colorlinks,linkcolor=red]{hyperref}
\usepackage{ulem}

\newcommand{\ee}{{\rm e}}
\def\tr{\ensuremath{\operatorname{tr}}}

\makeatletter
\let\cat@comma@active\@empty
\makeatother
\begin{document}
	
	\title{Topological Swing of Bloch Oscillations in Quantum Walks}
	
	\author{Lavi K. Upreti}
	
	\affiliation{Univ. Lyon, ENS de Lyon, Univ Claude Bernard, CNRS, Laboratoire de Physique, F-69342 Lyon, France}
	
	\author{C. Evain}
	\author{S. Randoux}
	\author{P. Suret}
	\author{A. Amo}
	\affiliation{Univ. Lille, CNRS, UMR 8523 -- PhLAM -- Physique des Lasers Atomes et Mol\'ecules, F-59000 Lille, France}
	\author{P. Delplace}
	\affiliation{Univ Lyon, ENS de Lyon, Univ Claude Bernard, CNRS, Laboratoire de Physique, F-69342 Lyon, France}

	\date{\today}
	
	\begin{abstract}
			We report new oscillations of wavepackets in quantum walks subjected to electric fields, that decorate the usual Bloch-Zener oscillations of insulators. The number of turning points (or sub-oscillations) within one Bloch period of these oscillations is found to be governed by the winding of the quasienergy spectrum. Thus, this provides a new physical manifestation of a topological property of periodically driven systems that can be probed experimentally. Our model, based on an oriented scattering network, is readily implementable in photonic and cold atomic setups. 
	\end{abstract}

	\maketitle
	

%
Periodically driven systems are routinely used to engineer artificial gauge fields and induce exotic topological properties. A remarkable case is the so-called anomalous topological boundary modes, which have no counterpart at equilibrium and were predicted \cite{Rudner2013} and observed in experimental platforms as different as photonics, cold atoms and acoustics~\cite{Rechtsman2013,Gao2016,Maczewsky2017,Mukherjee2017,Bellec2017,Afzal2020,Peng2016,Fleury2016,Wintersperger2020}. 
The existence of these states actually follows from the periodicity of the quasienergy spectrum that is itself inherited from the periodic drive.

Beyond these anomalous edge states, the frequency periodicity also allows the quasienergy spectrum to wind, which constitutes a distinct, so far largely overlooked topological property of the bands specific to periodically driven systems \cite{Tanaka2007,Miyamoto2007,Kitagawa2010}. One of the main consequences of this winding is the modification of the wavepacket dynamics in the bulk of the system.
The celebrated Thouless pumping, which consists of a quantized mean displacement of a particle in a 1D lattice subject to an adiabatic periodic drive \cite{Thouless1983}, was originally understood in terms of Berry curvature and was later rephrased as a winding property of the quasienergy spectrum with respect to the quasimomentum \cite{Kitagawa2010,Gong2016}.
Indeed, the measurement of wavepacket dynamics constitute one of the very few tools to probe geometrical and topological band properties of insulators. The associated anomalous velocity induces a measurable drift of the wavepacket that gives experimental access to the Berry curvature of the bulk bands \cite{Xiao2010}. This geometrical information was measured in periodically driven systems such as shaken trapped cold atoms gases \cite{Aidelsburger2014}, photonic quantum-walks \cite{Wimmer2017}, and in non-periodic structures \cite{Kraus2012}.

Here we report a new topological property of the motion of wavepackets that appears in periodically driven systems containing bands with a non trivial winding. While previous studies take advantage of usual Bloch oscillations to probe the lateral drifts associated to nontrivial Berry curvature~\cite{Longhi2007, Szameit2010, Cominotti2013, Flaschner2016, Li2016} or Berry-Zak phases of the bands~\cite{Atala2013, Holler2018, Maksimov2015, Kolovsky2018}, we find that periodically driven systems may exhibit a whole family of new Bloch-like oscillations with non-trivial properties. Differently from kicked rotors in which Bloch-like oscillations have been observed~\cite{Floss2014,Floss2015}, the oscillatory phenomena we unveil is directly related to the topology of the bands.
To explore this physics we propose a model of a 1D quantum walk with a time-varying synthetic electric field that is realisable in current photonic setups, and we find that the winding number characterising the bands governs the number of sub-oscillations of a wavepacket within a Bloch-Zener period. 
%
This topological feature guarantees the robustness of the number of these additional sub-oscillations irrespective of the coupling parameters of the model that, instead, determine their amplitude and shape. 
Finally, we discuss a parallel between these oscillations and Thouless pumping as they both reflect two complementary topological aspects of wavepackets dynamics that can be expressed in terms of distinct winding numbers of the quasienergy spectrum.
	
	

\begin{figure}[t!]
		\centering
		\includegraphics[width=0.7\columnwidth]{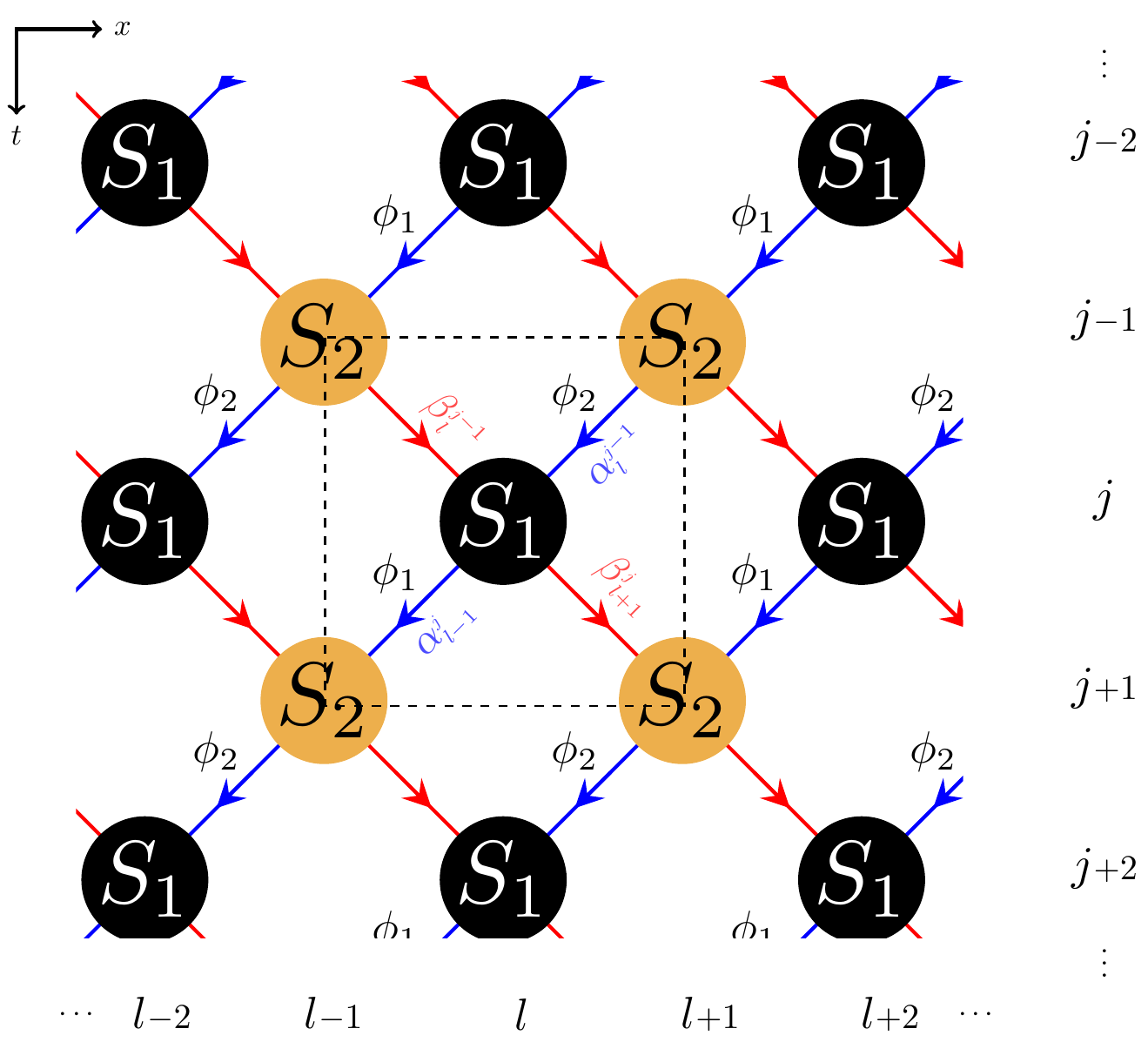}
		\caption{ Two-steps oriented scattering network, where an input signal flows from top to bottom.  
			Black and orange nodes represent two distinct scattering processes that repeat periodically. Two staggered phase shifts $\phi_1$ and $\phi_2$ are also considered along the leftward paths. A dashed square emphasizes the unit cell of this lattice.}
		\label{fig:sc4}
	\end{figure}

The model we consider is sketched in Fig.~\ref{fig:sc4}.  It consists of an oriented scattering network that represents a discrete-time evolution of a quantum state or a wavepacket through a periodic succession of scattering nodes (from top to bottom).
The discrete-time periodic dynamics are abundantly used in topological physics, both theoretically, where they were first introduced to illustrate the anomalous edge states \cite{Kitagawa2010, Rudner2013, Chong2013, Pasek2014, Tauber2015, Titum2016, Delplace2017, DelplaceGraph2020}, and experimentally in classical photonics \cite{HUPRX2015, Maczewsky2017, Mukherjee2017, Bellec2017}, in a quantum optics context \cite{Kitagawa2012, Derrico2018, Meng2019}, and in cold atomic setups \cite{Groh2016, Sajid2018}, where they are often referred to as discrete-time quantum walks \cite{KitagawaPRA2010, Asboth2015, Groh2016}.
The scattering network in Fig.~\ref{fig:sc4} is a representation of such 1D quantum walks. It has the advantage to be particularly suitable for a photonic interpretation, as the nodes simply describe beam-splitters and the oriented links represent the free propagation of a light pulse with time. It describes previous experiments, e.g., in 1D split-step quantum walks ~\cite{Kitagawa2012}, planar periodically coupled waveguides~\cite{Bellec2017} and coupled fiber rings~\cite{Wimmer2017, Weidemann2020}. In the following, we shall thus stick to this scattering network picture to model our periodic discrete-time evolution, but an equivalent Hamiltonian formalism is provided in Ref.~\onlinecite{Supplementary}.

The network depicts a succession of two scatterings events occurring at time steps $j$ and positions $l$, between incoming leftward and rightward wave amplitudes $\alpha_{l}^{j-1}$ and $\beta_{l}^{j-1}$, toward leftward and rightward outgoing states $\alpha_{l-1}^{j}$ and $\beta_{l+1}^{j}$. These scattering amplitudes are encoded in a dimensionless parameter $\theta_j$ entering the unitary matrix
	\begin{align}\label{eq:sj}
	S_{j} = \begin{pmatrix}
	\cos\theta_{j} & i\sin\theta_{j}\\ 
	i\sin\theta_{j}& \cos\theta_{j}
	\end{pmatrix} \ .
	\end{align}    

For simplicity, we consider only two distinct beam splitters $S_1$ and $S_2$ per time period, but the model can be generalised to further scattering events. Besides, we introduce a phase shift carried by the waves along their free propagation between these beam splitters, as in \cite{Wimmer2017}. Without any loss of generality, we consider a non-zero phase shift for the leftward states only (in blue in Fig.~\ref{fig:sc4}). 
	It was shown previously that such a phase induces a synthetic electric field when slowly varied in time over many periods \cite{Regensburger2012}. 
	A crucial point of the model is that this phase shift 
	takes two different values $\phi_1\neq \phi_2$ after each of the two scattering events ($S_1$ and $S_2$) within a period of the quantum walk. 
	The ratio of these phases is chosen to be a rational number, and we set $\phi_1=(m_1/n_1)\phi$ and $\phi_2=(m_2/n_2)\phi$ with  $ m_j, n_j \in \mathds{Z} $. As a result, the phase shift follows two time scales: a short one, that is the 2-step period of the photonic quantum walk (called Floquet period in the following), and which is fixed by the values of $m_j$ and $n_j$; and a much longer one over which $\phi$ may be adiabatically varied, in order to generate a fictitious electric field. The rapid variations of the phase within a Floquet period allow the generation of electric fields specific to periodically driven systems when $\phi$ is adiabatically varied. As we shall see, this gives rise to an original topological property of the evolution operator that manifests through unusual topological Bloch oscillations.

The time evolution of a state in the scattering network is given by relating the outgoing scattering amplitudes at time step $j$ to that at time step $j-1$, according to the scattering parameters Eq.\eqref{eq:sj}
	\begin{equation}
	\begin{aligned}\label{eq:timestep}
	\alpha_{l-1}^{j}&=(\cos \theta_{\tilde{j}} \, \alpha_{l}^{j-1}\,+ i \sin \theta_{\tilde{j}} \, \beta_{l}^{j-1}) \ee^{i \phi_{\tilde{j}}}\\
	\beta_{l+1}^{j}&=i \sin \theta_{\tilde{j}}\, \alpha_{l}^{j-1}\, + \cos \theta_{\tilde{j}} \, \beta_{l}^{j-1} \, .
	\end{aligned}   
	\end{equation}
where $ \tilde{j} = \text{mod}[j,2] $.
Assuming discrete translation invariance along the $x$-direction, the system can be treated in the Bloch-Floquet formalism. The corresponding Floquet unitary evolution operator $U_{F}(k,\phi)$ after a periodic sequence of $2$-steps readily describes a succession of local scattering events $S_j$ followed by rightward and leftward translations $T_j$:
\begin{align} \label{eq:floquet_def}
U_{F}(k,\phi_1,\phi_2) &= \ee^{i(\phi_{1}+\phi_{2})/2}\, T_2\,S_{2}\, T_1\, S_{1} \\
T_j& =\begin{pmatrix}
\ee^{i (k-\phi_{j}/2)} & 0\\
0 & \ee^{-i (k-\phi_{j}/2)}
\end{pmatrix}
\end{align}
where $k$ is the dimensionless quasimomentum in the $x$-direction, and we have removed the tilde from the $ j $ for the sake of clarity. Importantly, this quasimomentum is shifted by a phase $\phi_j$ that can thus be interpreted as a time-step dependent vector potential. 
Since the Floquet operator $U_F(k,\phi)$ depends periodically on the two variables $k$ and $\phi$, one can introduce a synthetic 2D Brillouin zone (BZ) to describe its eigenvalues spectrum. They decompose as $\lambda=\ee^{i \varepsilon}$, where $\varepsilon$ are hereafter referred to as the dimensionless quasienergy.

\begin{figure}[t!]
		\includegraphics[width=\columnwidth]{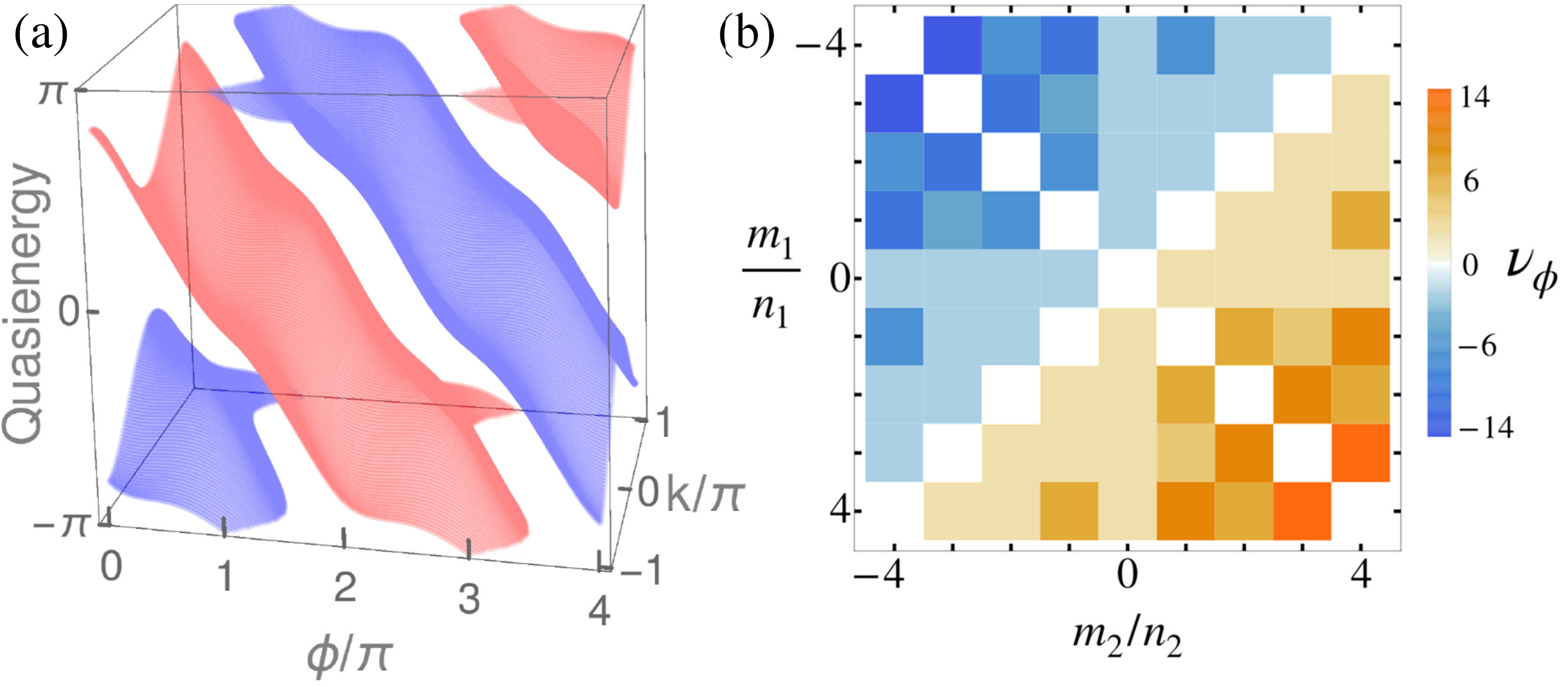}%
		\caption{(a)Quasienergy spectrum with a winding $ \nu_{\phi} = -2$ obtained for a scattering model with two-steps per period for $ \theta_{1} =\pi/4 $, $ \theta_{2} =\pi/4-0.6 $, $\phi_1=\phi$ and $\phi_2=-2\phi$. (b) Values of $\nu_\phi$ for integer values of $m_j/n_j$.}
		\label{fig:winding}
	\end{figure}

	The global phase factor in Eq.\eqref{eq:floquet_def} suggests that a peculiar spectral property arises when imposing a net phase  $\phi_1+\phi_2 \neq 0$ per period. 	In that case, the generalized inversion symmetry  $U_F(-k,-\phi) = \sigma_x U_F(k,\phi) \sigma_x$ (with $ \sigma_{x} $ the standard Pauli matrix) is broken~\cite{Supplementary}, leading to a winding of the quasienergy bands when  $\phi$ is varied, as illustrated in Fig.~\ref{fig:winding}(a).  A similar quasienergy winding was reported when considering periodically driven trapped cold atoms with a different protocol~\cite{Gong2016}.

	\begin{figure*}[t!]
	  \includegraphics[width=1.0\textwidth]{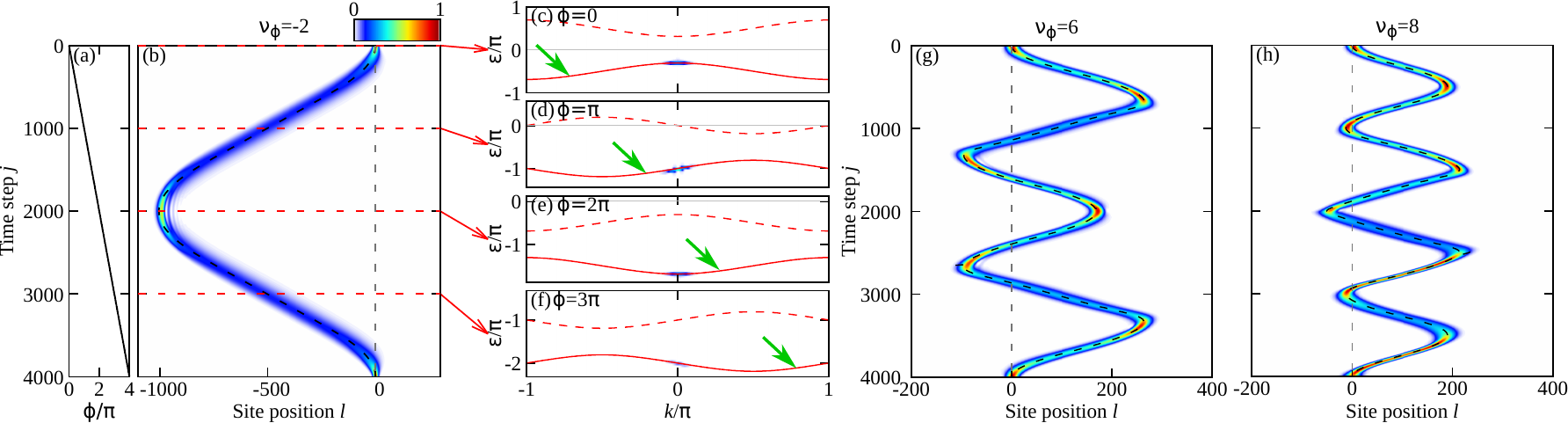}
         \caption{(a) Adiabatic increase of $\phi$ leads to (b) a standard Bloch oscillation ($\nu_\phi=-2$), and (g),(h) Bloch oscillation with sub-oscillations ($\nu_\phi=6$ and $\nu_\phi=8$ respectively) of a wavepacket injected at time $j=0$ and position $l=0$ with a Gaussian shape with rms width of 10 sites) in one band [the blue band in Fig.~\ref{fig:winding}(a) for the case of (b)]. Colorscale: intensity of the wavepacket ($|\alpha_l^j|^2+ |\beta_l^j|^2$). Dashed black line: analytical calculation of the centre of mass motion of a wavepacket from Eq.~(32) in~\cite{Supplementary}. (b),(g),(h) show one period $T_B$ of oscillation for the values of ($m_1$, $m_2$, $n_1$, $n_2$), $\theta_1$,  $\theta_2$ as (1,-2,1,1), $\pi/4$, $\pi/4-0.5$ for (b); (4,-1,1,1), $\pi/4$,$\pi/4-0.2$ for (g); and (9,-1,2,2), $\pi/4$,$\pi/4-0.2$ for (h). In (c)-(f), the norm of the 2D Fourier transform of the wavepacket ($\alpha$ part) after having evolved to the time step indicated by the horizontal lines in (b), and both the solid and the dashed red lines represent numerically calculated bands [Eq.~(33) in Ref.~\cite{Supplementary}]. The vertical scales differ in each panel. The green arrows show the direction in which the bands wind when $\phi$ increases.}
		\label{fig:osc}
	\end{figure*}

The period $\Phi$ of the BZ along $ \phi $ (i.e. $\varepsilon(k,\phi+\Phi)=\varepsilon(k,\phi)$) depends on $m_j$ and $n_j$  as $\Phi=4\pi\, \text{LCM}\left[(m_1/n_1 -m_2/n_2)^{-1},(m_1/n_1 + m_2/n_2)^{-1}\right]$, where LCM indicates the least common multiple~\cite{Supplementary}. 
	This allows us to define the winding of the quasienergies along $\phi$ as
	\begin{align}\label{eq:nuphi}
	\nu_{\phi} \equiv    \sum_{p=1}^N \frac{1}{2\pi }\int_{0}^{\Phi}  \dd\phi \frac{\partial \varepsilon_p(k,\phi)}{\partial \phi} = \frac{1}{2\pi i}\int_{0}^{\Phi}  \dd\phi \tr \Big[U_{F}^{-1}\partial_{\phi} U_{F}\Big] \ . 
	\end{align}
This winding number is a topological property of the Floquet evolution operator, as it reads as an element of the homotopy group $\pi_1[U(N)]=\mathbb{Z}$. 

	Note that $\nu_{\phi}$ is an even integer in our specific case due to the even number of bands ($ N = 2 $) in our model. A direct calculation leads to the simple result  
	\begin{align}
	\nu_{\phi} = \dfrac{\Phi}{2\pi} \left(\dfrac{m_1}{n_1} + \dfrac{m_2}{n_2}\right),
	\label{eq:winding}
	\end{align}
	which remarkably does not depend either on $k$ (since the winding of a quasienergy band $\varepsilon_p(k,\phi)$ along $\phi$ must be the same for any $k$) or on the scattering amplitudes $\theta_j$. Instead, it is proportional to the net phase $(\phi_1+\phi_2)/\phi$ that breaks inversion symmetry. A phase diagram representing the different possible values of $\nu_\phi$ as a function of $m_j/n_j$ is shown in Fig.~\ref{fig:winding}(b).

	A striking consequence of the winding of the quasienergy bands is the unconventional dynamics of the wavepackets in position space when adiabatically increasing the coordinate $\phi$. In the following, we show how these dynamics reveal a new kind of Bloch oscillations described by the winding number $\nu_\phi$. Figure~\ref{fig:osc}(b) shows the $j$-time evolution of a Gaussian wavepacket injected at $j=0$ in the blue band of Fig.~\ref{fig:winding}(a) at $k=0$, when $\phi$ is adiabatically increased from 0 to $\Phi=4 \pi$, with $\phi(j)=\gamma_0 j$ where the rate $\gamma_0=2\pi/2000$ [see Fig.~\ref{fig:osc}(a)]. To compute the spatio-temporal dynamics, we apply Eq.~\eqref{eq:timestep} to the initial wavepacket. The wavepacket periodically oscillates in space coordinate while keeping $k$ constant. This can be readily seen in Fig.~\ref{fig:osc}(c)-(f), where we show the 2D Fourier transform of the spatio-temporal dynamics of the wavepacket after having evolved to the time step indicated by the horizontal lines in Fig.~\ref{fig:osc}(b). These panels provide a phenomenological understanding of the mechanism behind the oscillations: as $\phi$ is adiabatically increased, the band dispersions are displaced in a diagonal direction in $(k,\varepsilon)$ space [green arrows in Fig.~\ref{fig:osc}(c)-(f)], a direct consequence of the winding of the bands [see also Fig.~\ref{fig:winding} (a)]. Therefore, the group velocity $v_g=\frac{\partial\varepsilon}{\partial k}$ of a wavepacket with a given $k$ changes sign when $\phi(j)$ increases, resulting in oscillations in the spatial coordinate. 
	
It is worth stressing that two distinct drivings are present in our model: (i) a fast cyclic driving of the phases $\phi_1$, $\phi_2$ within a Floquet period, which confers a nontrivial winding to the bands; (ii) a slow adiabatic increase of the phase $\phi$ which results in the oscillations.
	An analytical calculation of the center of mass trajectory $X_c(t,k)$ of the wavepacket initially injected at a given $k$ can be inferred from the group velocity of the quasienergy bands in parameter space~\cite{Supplementary}:
	\begin{align}
	X_c(t,k) = \gamma_0\int_0^t \dd \tau\,  v_g(\phi(\tau),k), 
	\label{eq:classique}
	\end{align}
	where the continuous-time variable $t$ extrapolates the discrete one $j$. This semiclassical trajectory is shown in black dashed lines in Fig.~\ref{fig:osc}(b), which fits the simulation plot perfectly.
	
	More importantly, the observed oscillatory phenomenon establishes a direct connection between the winding of the bands in our Floquet-Bloch model and the usual Bloch oscillations in a periodic crystal subject to a constant electric field. Indeed, the adiabatic increase of the phase shift $\phi$ at a rate $\gamma_0$ when the time steps $j$ increase is analogous to a time-dependent vector potential that induces a (fictitious) electric field~\cite{Krieger1986} $E$ and, therefore, should result in Bloch oscillations. This was already noticed in the case of a single-step time evolution by Wimmer and co-workers~\cite{Wimmer2015}, who reported a gauge transformation relating the dynamics of a wavepacket in a lattice subject to a static potential gradient (i.e., a constant electric field), and the dynamics in a lattice subject to an adiabatic increase of the parameter $\phi$ (see also~\cite{Supplementary}).
	
	To establish the connection between the winding of the quasienergy bands and Bloch oscillations, we note that the time periodicity $T_B$ of the center of mass motion $X_c$ is inherited from the periodicity of the quasienergy with respect to $\phi$ in the following way
	\begin{align}
	T_B = \frac{2\pi}{\gamma_0} \frac{\nu_\phi}{\frac{m_1}{n_1}+\frac{m_2}{n_2}}, 
	\label{eq:period}
	\end{align} 
	where negative values of $\nu_\phi$ correspond to mirror symmetric trajectories to those with $|\nu_\phi|$.
	
	In Eq.~\eqref{eq:period}, we recognize the usual period $T_B$ for Bloch oscillations induced by an average constant electric field $E=(E_1+E_2)/2$ where $E_j=\frac{m_j}{n_j}\gamma_0$ is the fictitious electric field applied during the time step $j$ (see Ref.~\cite{Supplementary} for more details), except that in Eq.~\eqref{eq:period} this standard relation is modified by the winding number $\nu_\phi$. In particular, the period $T_B=2\pi/ E$ of the usual Bloch oscillations is recovered for $\left|\nu_\phi\right|=2$, a situation in which each band winds once, as reported in Figs.~\ref{fig:winding}(a) and~\ref{fig:osc}(b).
	
	Beyond this standard case, our model predicts a novel kind of topological oscillations: higher winding numbers may not only change the period $T_B$, but also yield more complex oscillations with additional turning points within $T_B$. Two examples are shown in Fig.~\ref{fig:osc}(g-h) for values of $m_i$, $n_i$ resulting in bands of windings  $\nu_\phi=6$ and~$8$, respectively, and same oscillating period $T_B$ as in Fig.~\ref{fig:osc}(b). Remarkably, in a period $T_B$, the number of turning points is found to be precisely $\mathcal{N}_{t}=|\nu_\phi |$ (see Ref.~\cite{Supplementary}). This result confers a topological nature to the observed oscillations. Note that standard Bloch oscillations simply have two turning points per period (see Fig.~\ref{fig:osc}(b)), in agreement with $\mathcal{N}_t=2=|\nu_\phi|$. Moreover, the topological nature of the oscillations makes them robust to the presence of weak disorder in the lattice~\cite{Supplementary}.

	
	\begin{figure}[t]
		\centering
		\includegraphics[width=0.48\textwidth]{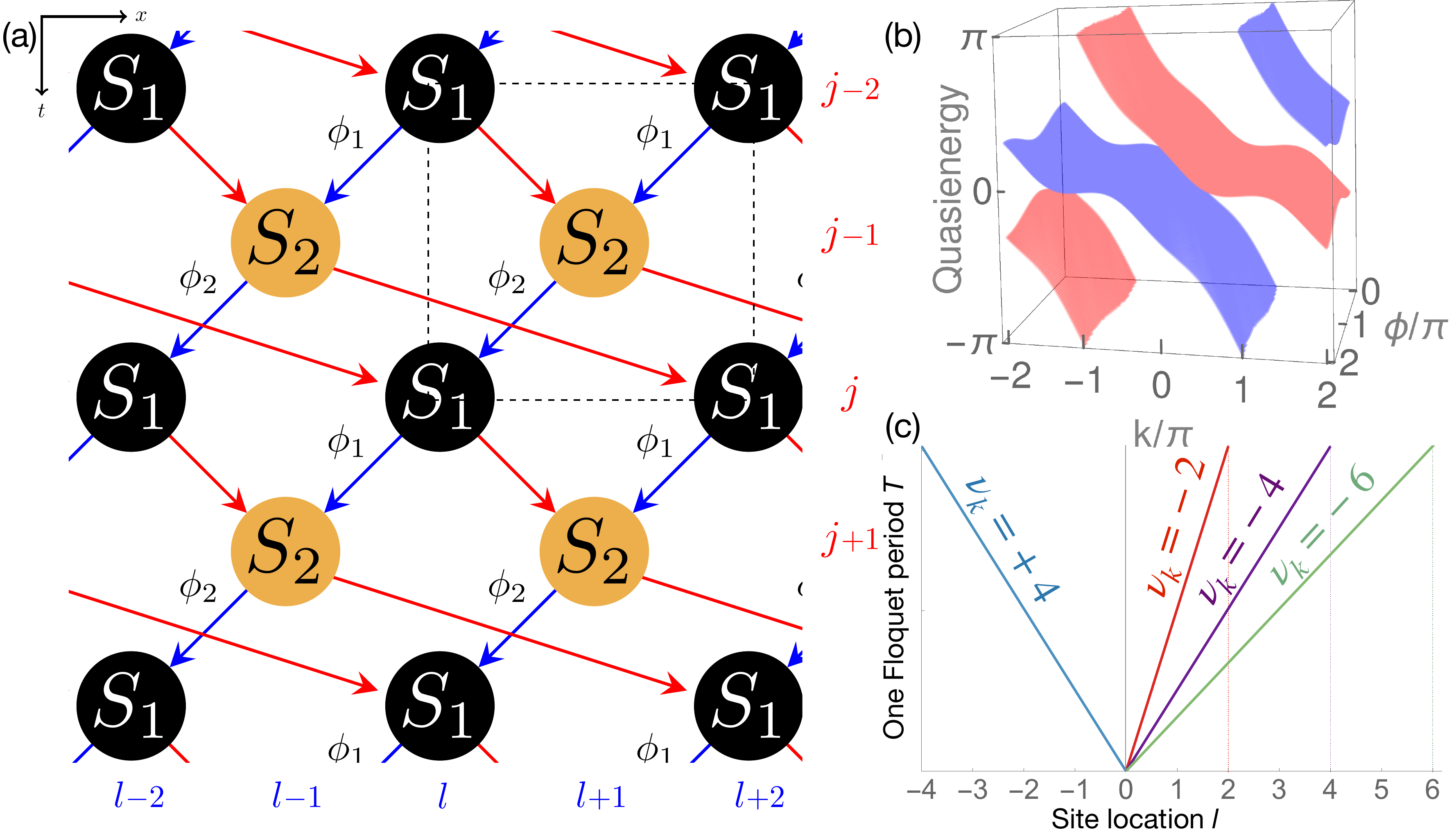}
		\caption{(a) Two-steps scattering network with the next nearest coupling in the second step. A dashed black rectangle emphasizes the unit cell of this lattice. (b) Quasienergy bulk spectrum for the model depicted in (a) with $\theta_1=\pi/4$, $\theta_2=\pi/4-0.6$ and $\phi_1=-\phi_2$. (c) Quantized displacement of the mean particle position with associated winding numbers $\nu_k$.}
		\label{fig:sck}
	\end{figure}

	So far, we have considered windings of the bands induced by periodic pumping in the synthetic dimension. We now show that a winding of the quasienergy bands along the $k$-direction can similarly be induced when inversion symmetry is broken in the spatial dimension, and it results in a different topological phenomenon: quantized displacement of the mean particle position. This effect can be straightforwardly implemented in scattering network models by connecting next-nearest neighbor nodes, as sketched in Fig.~\ref{fig:sck}(a) for a two-step time evolution (see Ref.~\cite{Supplementary} for the step evolution equations). For the sake of generality, we have kept in the model the phase $\phi_{j}$, which we take as $\phi_1=\phi=-\phi_2$, such that $\nu_\phi=0$, i.e., there are no Bloch oscillations. The corresponding quasienergy bands, displayed in Fig.~\ref{fig:sck}(b), show a winding along $k$ for each $\phi$.
	This feature is captured by a winding number of the Floquet operator along $k$, analogous to that defined in Eqs.~\eqref{eq:nuphi}-\eqref{eq:winding} for $\phi$. More generally, when considering even further long range couplings, this winding number is found to read (\cite{Supplementary}):
	\begin{align}
	\nu_{k} = \dfrac{\kappa}{2\pi} \left(\dfrac{r_1}{s_1} + \dfrac{r_2}{s_2}\right) 
	\label{eq:windingk}
	\end{align}
where $\kappa$ is the periodicity of the bands in $k$ and ${r_j}/{s_j}$ is related to the range of the couplings between nodes to the left or to the right at each time step $j$. For the case illustrated in Fig.~\ref{fig:sck}(a), ${r_1}/{s_1}=1$, ${r_2}/{s_2}=-2$.
	
	In the spirit of the seminal work of Thouless~\cite{Thouless1983}, and as revisited by Kitagawa \textit{et al.} \cite{Kitagawa2010} within the Floquet formalism, this winding number $\nu_{k}$ can be related to the mean displacement of particles after $P$ Floquet periods $T$, in a state where  all the bands are uniformly excited, that is \cite{Supplementary}, which also includes~\cite{Zak1988}, $\Delta x  =-P \dfrac{2\pi}{\kappa}\nu_k$. 

	Despite this apparent similarity, this quantized transport property differs from the usual Thouless pumping that results from an \textit{adiabatic} driving of the system. In that case, the quantization  can be expressed as a Chern number of the slowly driven instantaneous filled states parametrized over the effective 2D BZ $(k,t)$. This Chern number was later reinterpreted  as a sum of the winding numbers in $k$ over the filled bands \cite{Kitagawa2010}. In the adiabatic regime, if this sum runs over all the bands, as in our case, then the Chern numbers of each band sum up to zero, and there is no drift. Quantized drifts obtained for our \textit{non-adiabatic} scattering model are shown in Fig.~\ref{fig:sck}(c). More generally, quasienergy windings along both $\phi$ and $k$ coordinates can coexist, leading to quite complex drifted Bloch oscillations for wavepackets, which are shown in~\cite{Supplementary}.

Our study unveils the topological aspects of Bloch oscillations and extends them to a family of oscillatory phenomena accessible in artificial systems such as arrays of photonic waveguides and coupled fibers. It generalizes straightforwardly to periodically driven lattices of ultracold atoms where a protocol to generate quasienergy windings and oscillations was proposed \cite{Gong2016}, although neither the winding number $\nu_\phi$ nor its relation to the number of Bloch sub-oscillations was identified. The direct relation we have identified between the number of turning points within an oscillation period and the winding number of the bands provides a new protocol to measure topological invariants in systems described by a quantum walk. The study of the manifestation of the topological nature of the oscillations in their associated Wannier-Stark ladders~\cite{Mendez1988} remains an exciting perspective.

\textit{Acknowledgements}-- The authors are thankful to Benoit Dou\c cot for his very constructive comments. This work was supported by the French Agence Nationale de la Recherche (ANR) under grant Topo-Dyn (ANR-14-ACHN-0031), the Labex CEMPI (ANR-11-LABX-0007), the CPER Photonics for Society P4S, the I-Site ULNE project NONTOP, the \textit{M\'etropole Europ\'eenne de Lille} (project TFlight) and the European Research Council (ERC) under the European Union’s Horizon 2020 research and innovation programme (Grant Agreement No.~865151).

	\bibliography{Windingbib}

\onecolumngrid

\section{S1. Scattering network model}
\begin{figure}[!htb]
	\centering
	\includegraphics[width=0.7\linewidth]{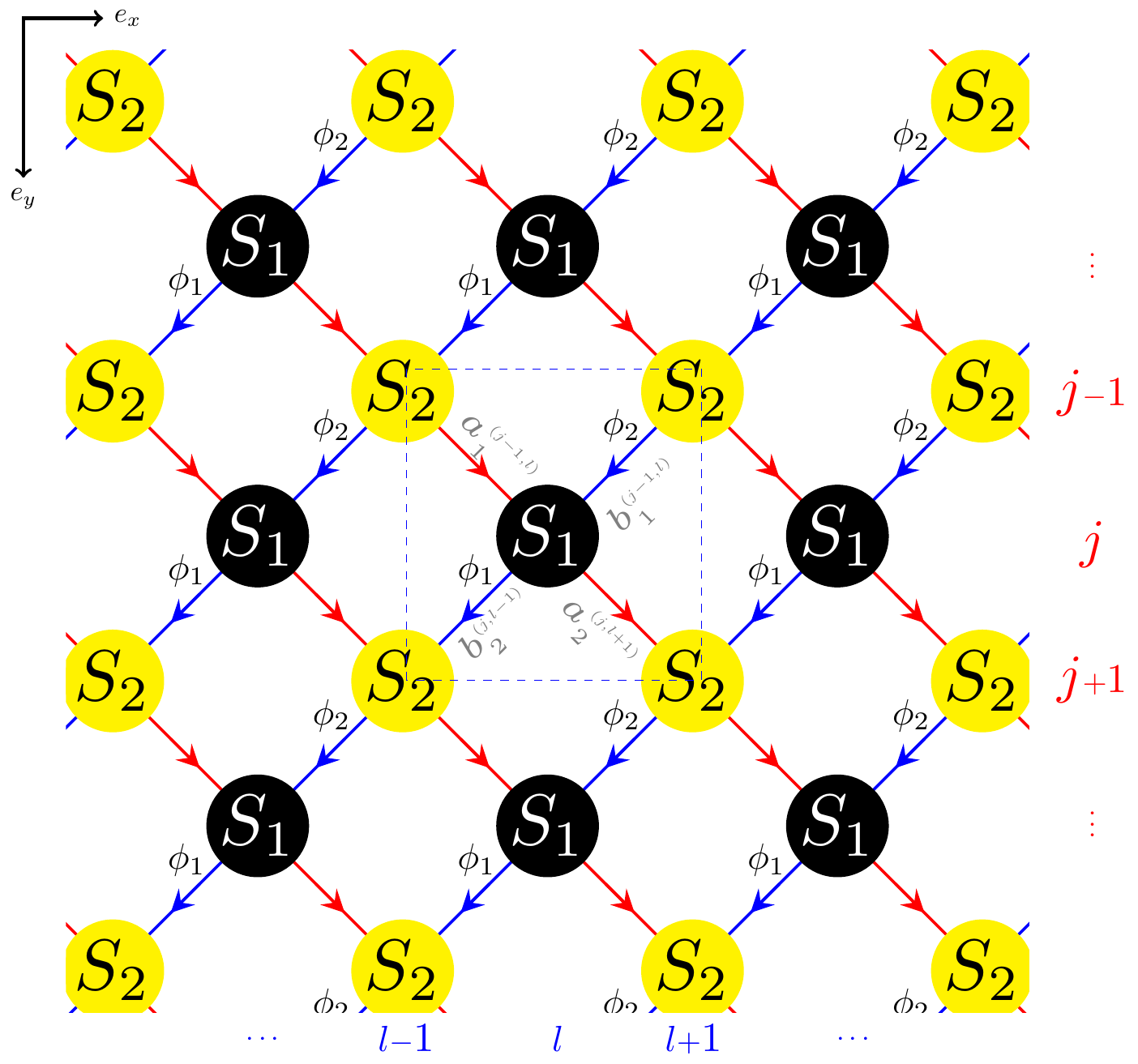}
	\caption{2D oriented two-steps scattering network model with a preferential direction from top to bottom.}
	\label{fig:scatnet1}
\end{figure}

We consider the 2D oriented scattering network defined in the main text, and reproduced in Fig.~\ref{fig:scatnet1} for the time period of two time-steps. We detail in this section the derivation of the evolution operator, its quasienergies and the center of mass trajectories showing Bloch oscillations. 

\subsection{Derivation of the Floquet evolution operator}

The oriented network shown in Fig.~\ref{fig:scatnet1} is constituted of two distinct successive scattering nodes $ S_{1} $ and $ S_{2} $. The incoming arrow from left (right) toward the $ S_{1}$ node is denoted by $ a_{1} $ $ (b_{1}) $. It denotes a time evolution from the time step $ j-1 $ to time step $ j $. Similarly, the outgoing arrows are denoted by $ a_{2}, b_{2} $. These four oriented paths and the two scattering nodes constitute the unit cell of the network, which is emphasized with a dashed square in Fig.~\ref{fig:scatnet1}. The dynamics is then given by the relations
\begin{equation}\label{eq:s1}
	\left(\begin{array}{c}
		a_{2}(j,l+1) \\ 
		b_{2}(j,l-1)
	\end{array}\right) = S_{1} \left(\begin{array}{c}
		a_{1}(j-1,l) \\ 
		b_{1}(j-1,l)
	\end{array}\right)
\end{equation}
and
\begin{equation}\label{eq:s2}
	\left(\begin{array}{c}
		a_{1}(j-1,l) \\ 
		b_{1}(j-1,l)
	\end{array}\right) = S_{2} \left(\begin{array}{c}
		a_{2}(j-2,l+1) \\ 
		b_{2}(j-2,l-1)
	\end{array}\right) \ ,
\end{equation}
which can be grouped together as
\begin{equation}\label{eq:s12}
	\left(\begin{array}{c}
		a_{1}(j-1,l) \\ 
		b_{1}(j-1,l) \\
		a_{2}(j,l+1) \\ 
		b_{2}(j,l-1)
	\end{array}\right) = \left(\begin{array}{cc}
		0 & S_{2} \\ 
		S_{1} & 0
	\end{array}\right)  \left(\begin{array}{c}
		a_{1}(j-1,l) \\ 
		b_{1}(j-1,l) \\
		a_{2}(j-2,l+1) \\ 
		b_{2}(j-2,l-1)
	\end{array}\right) \ .
\end{equation} 
The current scattering matrix notations correspond to that of the main text (Eq.(2)), as follows $ a_{1}(j-1,l) = \alpha_{l}^{j-1}$, $ b_{1}(j-1,l) = \beta_{l}^{j-1}$, similarly, $ a_{2}(j,l+1) = \alpha_{l+1}^{j}$, $ b_{2}(j,l-1) = \beta_{l-1}^{j}$.\\

Using translation symmetry of the scattering network, we can Fourier decompose as
\begin{equation}\label{eq:ft}
	\left(\begin{array}{c}
		a_{m}(j,l) \\ 
		b_{m}(j,l)
	\end{array}\right) = \sum\limits_{k_{x},k_{y}} \ee^{i \vec{k}.(l\hat{e}_{x}+j\hat{e}_{y})/2} \left(\begin{array}{c}
		a_{m}(k_{x},k_{y}) \\ 
		b_{m}(k_{x},k_{y})
	\end{array}\right), \quad  m=1,2 \ .
\end{equation}
This gives,
\begin{eqnarray}\label{eq:ufl}
	\begin{pmatrix}
		a_{1}(k_{x},k_{y}) \\
		b_{1}(k_{x},k_{y}) \\
		a_{2}(k_{x},k_{y}) \\
		b_{2}(k_{x},k_{y})
	\end{pmatrix} & =&\begin{pmatrix}
		0 & 0 & s_{2}^{11}\ee^{i k_{x}/2}\ee^{-i k_{y}/2}  & s_{2}^{12}\ee^{-i k_{x}/2}\ee^{-i k_{y}/2} \\ 
		0 & 0 & s_{2}^{21}\ee^{i k_{x}/2}\ee^{-i k_{y}/2} & s_{2}^{22}\ee^{-i k_{x}/2}\ee^{-i k_{y}/2} \\ 
		s_{1}^{11}\ee^{i k_{x}/2}\ee^{-i k_{y}/2} & s_{1}^{12}\ee^{-i k_{x}/2}\ee^{-i k_{y}/2} & 0 & 0 \\ 
		s_{1}^{21}\ee^{i k_{x}/2}\ee^{-i k_{y}/2} & s_{1}^{22}\ee^{-i k_{x}/2}\ee^{-i k_{y}/2} & 0 & 0
	\end{pmatrix}\begin{pmatrix}
		a_{1}(k_{x},k_{y}) \\ 
		b_{1}(k_{x},k_{y}) \\
		a_{2}(k_{x},k_{y}) \\ 
		b_{2}(k_{x},k_{y})
	\end{pmatrix}\\
	\label{eq:ufl1}\begin{pmatrix}
		\vec{a_{1}}(\vec{k})\\
		\vec{a_{2}}(\vec{k}) 
	\end{pmatrix}&=& \begin{pmatrix}
		0 & \mathcal{S}_{2}(\vec{k}) \\ 
		\mathcal{S}_{1}(\vec{k}) & 0
	\end{pmatrix}
	\begin{pmatrix}
		\vec{a_{1}}(\vec{k})\\
		\vec{a_{2}}(\vec{k}) 
	\end{pmatrix}
\end{eqnarray}
where $ \vec{a_{1}}(\vec{k}) = \{a_{1}(\vec{k}), b_{1}(\vec{k})\} $ and $ s_{j}^{m_1 m_2} $ ($ m_1,m_2 = 1,2 $), are the scattering coefficients of the scattering matrix $ S_{j} $. In the main text, we choose 
\begin{equation}\label{eq:scon}
	S_{j} = \left(\begin{array}{cc}
		\cos\theta_{j} & i\sin\theta_{j} \\ 
		i\sin\theta_{j}& \cos\theta_{j}
	\end{array} \right).
\end{equation}
although the calculations are independent of this specific form.
Squaring Eq.\eqref{eq:ufl1} allows one to define the Floquet operators starting for different time origins as
\begin{equation}
	\left(
	\begin{array}{cc}
		\mathcal{S}_{2}(\vec{k}) \mathcal{S}_{1}(\vec{k}) & 0 \\
		0 & \mathcal{S}_{1}(\vec{k}) \mathcal{S}_{2}(\vec{k})
	\end{array}
	\right)
	=   
	\begin{pmatrix}
		U_F^{21}(\vec{k}) & 0 \\
		0 &       U_F^{12}(\vec{k})
	\end{pmatrix}
\end{equation}
Substituting Eq.\eqref{eq:scon} gives more specifically
\begin{eqnarray}\label{eq:s2s1}
	U_F^{21}(\vec{k})&=&\left(
	\begin{array}{cc}
		\ee^{-i k_y}(\ee^{i k_x} \cos\theta _1\cos\theta _2-\sin\theta _1 \sin\theta _2) & i \ee^{-i k_y}(\cos\theta _2\sin\theta _1+\ee^{-i k_x}\cos \theta _1 \sin \theta _2)\\
		i \ee^{-i k_y}(\cos\theta _2\sin \theta _1+\ee^{i k_x}\cos \theta _1 \sin \theta _2) & \ee^{-i k_y}(\ee^{-i k_x} \cos \theta _1 \cos \theta _2-\sin \theta _1 \sin \theta _2)
	\end{array}
	\right) \\
	\label{eq:s1s2}
	U_F^{12}(\vec{k})&=&\left(
	\begin{array}{cc}
		\ee^{-i k_y}(\ee^{i k_x} \cos \theta _1 \cos \theta _2-\sin \theta _1 \sin \theta _2) & i \ee^{-i k_y}(\ee^{-i k_x}\cos \theta _2 \sin \theta _1+\cos \theta _1 \sin \theta _2) \\
		i \ee^{-i k_y}(\ee^{i k_x}\cos \theta _2 \sin \theta _1+\cos \theta _1 \sin \theta _2) & \ee^{-i k_y}(\ee^{-i k_x} \cos \theta _1 \cos \theta _2-\sin \theta _1 \sin \theta _2)
	\end{array}
	\right)
\end{eqnarray}
Then, we add a phase $ \phi $ to the $ b_{j} $ amplitudes, that is to the blue arrows in Fig. \eqref{fig:scatnet1}, such that $ b_{1}\rightarrow b_{1} \ee^{i\phi_{2}} $ and $ b_{2}\rightarrow b_{2} \ee^{i\phi_{1}} $. Then in Eq.\eqref{eq:ufl}, $ s_{2}^{12} $ and  $ s_{2}^{22} $ will be multiplied by $ \ee^{i\phi_{1}} $, likewise, $ s_{1}^{12} $ and  $ s_{1}^{22} $ are multiplied by $ \ee^{i\phi_{2}} $. That gives
\begin{eqnarray}\label{eq:s2s1f}
	U_F^{21}(\vec{k},\phi)&=&\ee^{-i k_y}\left(
	\begin{array}{cc}
		\ee^{i k_x} \cos\theta _1\cos\theta _2-\ee^{i \phi_{1}}\sin\theta _1 \sin\theta _2 & i (\ee^{i \phi_{2}}\cos\theta _2\sin\theta _1+\ee^{-i k_x}\cos \theta _1 \sin \theta _2)\\
		i(\ee^{i \phi_{1}}\cos\theta _2\sin \theta _1+\ee^{i k_x}\cos \theta _1 \sin \theta _2) & \ee^{-i k_x} \cos \theta _1 \cos \theta _2-\ee^{i \phi_{2}}\sin \theta _1 \sin \theta _2
	\end{array}
	\right) \\
	\label{eq:s1s2f}
	U_F^{12}(\vec{k},\phi)&=&\ee^{-i k_y}\left(
	\begin{array}{cc}
		\ee^{i k_x} \cos \theta _1 \cos \theta _2-\ee^{i \phi_{2}}\sin \theta _1 \sin \theta _2 & i (\ee^{-i k_x}\cos \theta _2 \sin \theta _1+\ee^{i \phi_{1}}\cos \theta _1 \sin \theta _2) \\
		i(\ee^{i k_x}\cos \theta _2 \sin \theta _1+\ee^{i \phi_{2}}\cos \theta _1 \sin \theta _2) & \ee^{-i k_x} \cos \theta _1 \cos \theta _2-\ee^{i \phi_{1}}\sin \theta _1 \sin \theta _2
	\end{array}
	\right).
\end{eqnarray}
These two evolution operators describe the same physical system, and either of them can be used to compute the quasienergy spectrum and the winding numbers.  The common phase factor $\exp{-i k_y}$ in Eqs.\eqref{eq:s2s1f}-\eqref{eq:s1s2f}  is reminiscent of the preferential orientation of the network from top to bottom. This is the only $k_y$ dependence of the evolution operator on the network. In the main text, the Floquet operator refers to $U_F^{21}$ where this phase factor is factorized out, that is 
\begin{align}
	U_F(k_x,\phi)\equiv U_F^{21} (k,\phi) \ee^{i k_y} 
\end{align}
and we set $k_x=k$ through out the paper. The eigenvalues of $U_F$ are defined as $\ee^{-i \epsilon T}\equiv \ee^{i \varepsilon}$, where the dimensionless quasienergy $\varepsilon$ is the quantity considered in the main text. Then the Floquet operator can usefully be factorized as 
\begin{eqnarray}\label{eq:UF}
	U_{F} =  B_0(k) S_2D(\phi_2) B_1(k) S_1D(\phi_1),
\end{eqnarray}
where \begin{equation}\label{eq:sbdmatrix}
	B_1(k)=\begin{pmatrix}
		1 & 0\\
		0 & \ee^{-i k}
	\end{pmatrix},
	\  B_0=\begin{pmatrix}
		\ee^{i k} & 0\\
		0 & 1
	\end{pmatrix},
	\ D_{j}= D(\phi_j) =\begin{pmatrix}
		1 & 0\\
		0 & \ee^{i \phi_{j}}
	\end{pmatrix}.
\end{equation}

\subsection{Alternative Hamiltonian formalism }
\begin{figure}[h!]
	\includegraphics[width=0.7\columnwidth]{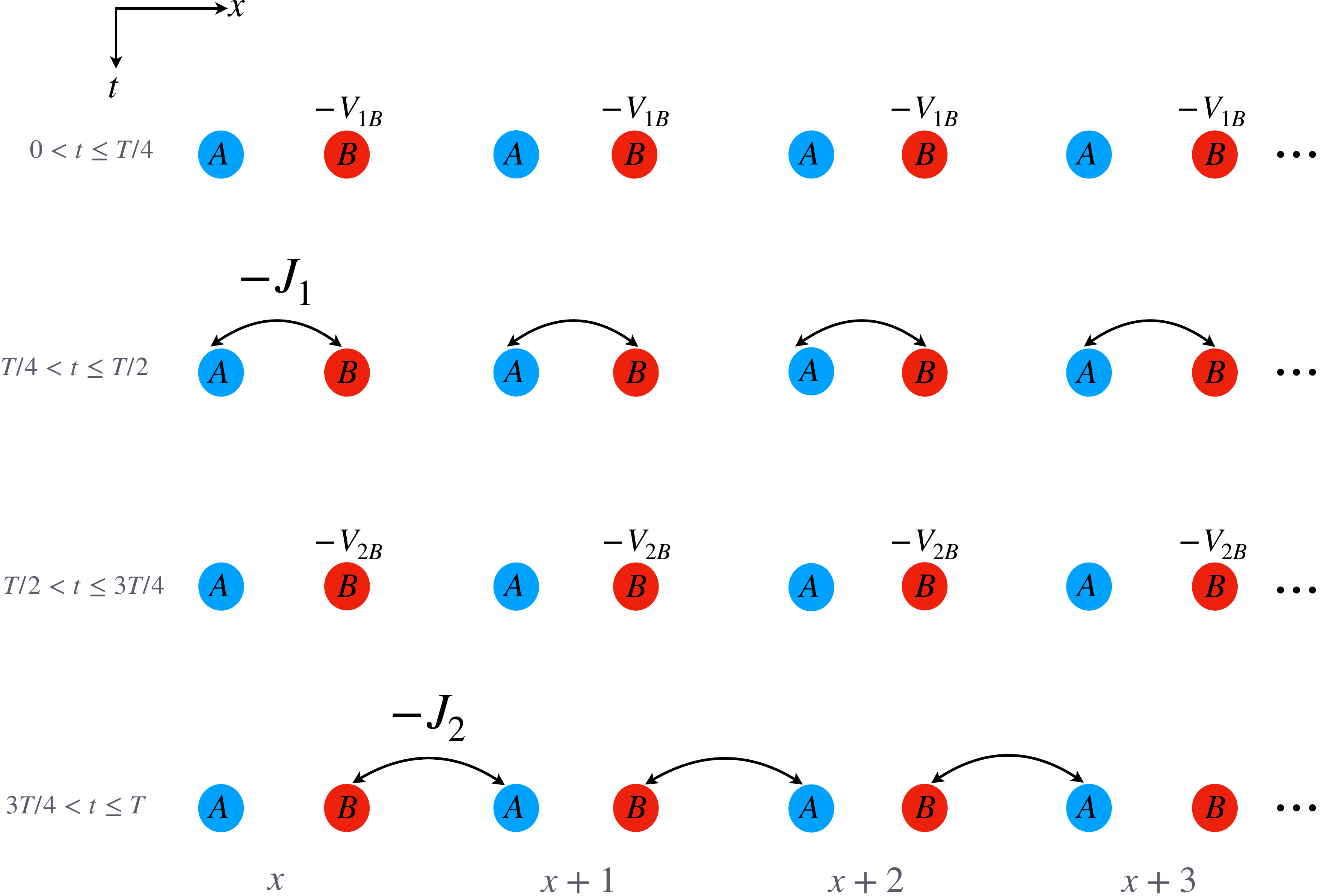}
	\caption{Periodically driven lattice model whose Floquet operator corresponds to that of the scattering network of the main text. The driving period is made of four steps during which either an onsite potential is switched on onto one sublattice (steps 1 and 3), or a hopping term is switched to dimerized the lattice (steps 2 and 4). The lattice spacing $a$ is set to $1$. }
	\label{fig:lattice_ham_winding}
\end{figure}

We propose here an equivalent Hamiltonian formalism to the scattering model discussed in the main text and the previous section. An oriented network consisting of two scattering processes with two free propagations per period can be mapped onto a four time-step periodically driven tight-binding Hamiltonian. The scattering parameters of the network are related to the nearest-neighbors hopping terms $J_i$ of a driven lattice, while the phase shifts accumulated during the free propagations are related to on-site potential $J_i$. The corresponding periodically driven 1D lattice over a time period $T$ is depicted in Fig.~\ref{fig:lattice_ham_winding} in the case of two-steps, as detailed in the main text. It is composed of two atoms per unitcell, denoted by $ A $ and $ B $ (similar to the two arrows, red and blue, entering each scattering node in Fig.~1 of main text), and is driven over $4$ steps so that the stepwise Bloch Hamiltonian reads

\begin{align}
	H(t,k_x) =
	\left\{
	\begin{array}{ll}
		H_1(k_x) = \begin{pmatrix}
			0 & 0 \\
			0  & -V_{1}
		\end{pmatrix}, \quad &0<t\le t_1\\
		H_2(k_x) =  \begin{pmatrix}
			0 & -J_{1}  \text{e}^{-i k_{x}/2}\\
			-J_{1}  \text{e}^{i k_{x}/2} & 0
		\end{pmatrix}\quad &t_1<t\le t_2\\
		H_3(k_x) = \begin{pmatrix}
			0 & 0 \\
			0  & -V_{2}
		\end{pmatrix} \quad &t_2<t\le t_3\\
		H_4(k_x) = \begin{pmatrix}
			0 & -J_{2} \text{e}^{i k_{x}/2} \\
			-J_{2} \text{e}^{-i k_{x}/2}  & 0
		\end{pmatrix}  \quad  &t_3<t\le T
	\end{array}
	\right.
\end{align}

Accordingly, a stepwise evolution operator $U_j$ during the duration $\tau_j=t_j-t_{j-1}$ can be defined as 
\begin{align}\label{eq:wind_hamiltonian2}
	U_{j}(k_x) \equiv \text{e}^{-i H_{j}(k_x) \tau_j/\hbar},
\end{align}
so that the evolution (Floquet) operator after one period is defined as  $ U_{F}(k_x) = U_{4} U_{3} U_{2} U_{1} $.  By setting 
\begin{align}
	\phi_{1} \equiv V_{1} \tau_1/\hbar \quad \quad \theta_{1} \equiv J_{1}\tau_2/\hbar  \quad \phi_{2} \equiv V_{2} \tau_3/\hbar \quad \theta_{2} \equiv J_{2}\tau_4/\hbar 
\end{align}
the stepwise evolution operators read
\begin{align}\label{eq:U1matrix}
	U_1=  \begin{pmatrix}
		1 & 0 \\
		0  & \text{e}^{i \phi_{1}}
	\end{pmatrix}
	\qquad U_2=   \begin{pmatrix}
		\cos \theta_{1} & i \ee^{-i k_{x}/2}\sin \theta_{1} \\
		i \ee^{i k_{x}/2}\sin \theta_{1}  & \cos \theta_{1}
	\end{pmatrix}\\
	U_3=  \begin{pmatrix}
		1 & 0 \\
		0  & \text{e}^{i \phi_{2}}
	\end{pmatrix}
	\qquad U_4=   \begin{pmatrix}
		\cos \theta_{2} & i \ee^{-i k_{x}/2}\sin \theta_{2} \\
		i \ee^{i k_{x}/2}\sin \theta_{2}  & \cos \theta_{2}
	\end{pmatrix}
\end{align}
leading to the expression of the Floquet operator, as given in the main text (Eq.(3)).

\subsection{Generalized inversion symmetry breaking } \label{sub:inv}
Considering the two parameters $\phi$ and $k$ on the same footing allows us to define the generalized inversion symmetry $U_F(k,\phi) = \sigma_x U_F(-k,-\phi) \sigma_x $, where $\sigma_x$ is the standard Pauli matrix.
The existence of a net phase in the unit cell, (i.e. $\phi_1+\phi_2\neq 0$, with $\phi_1$ and $\phi_2$ proportional to $\phi$) breaks this symmetry.
This can be shown by symmetrizing the $B_j$ matrices in Eq. \eqref{eq:UF} as
\begin{eqnarray}
	U_{F}(k,\phi) &=& B_{0}(k)\,  S_{2} \, D(\phi_2) \, B_{1}(k)\,  S_{1} \, D(\phi_1)
	= B(k)\,  S_{2} \, D(\phi_2) \, B(k)\,  S_{1} \, D(\phi_1)
\end{eqnarray}
where,
\begin{equation}\label{eq:bk}
	B(k) = \begin{pmatrix}
		\ee^{i k/2}  & 0 \\
		0 & \ee^{-i k/2}
	\end{pmatrix}
\end{equation}
and doing so as well for the $D_j$ matrices, thus factorizing the net phase
\begin{align} \label{eq:UF_net}
	U_{F}(k,\phi) = \ee^{i(\phi_{1}+\phi_{2})/2}  B(k)\,  S_2\,  \tilde{D}(\phi_2)\,  B(k)\,  S_1 \, \tilde{D}(\phi_1)
\end{align}
(as already defined as $ U_{1,3} $ in Eq.\eqref{eq:U1matrix} from Hamiltonian picture),
\begin{equation}\label{eq:dphi}
	\tilde{D}(\phi_{j})=\begin{pmatrix}
		\ee^{-i \phi_j/2} & 0\\
		0 & \ee^{i\phi_j/2}
	\end{pmatrix}
\end{equation}
Next we notice that $\sigma_x B(k) \sigma_x = B(-k)$ and $\sigma_x \tilde{D}(\phi_{j}) \sigma_x = \tilde{D}(-\phi_{j})$ where we recall that $\phi_j$ is proportional to $\phi$. Therefore, the net phase, in the phase factor in Eq. \eqref{eq:UF_net} prevents $U_F$ to be inversion symmetric
that is
\begin{align}
	\sigma_x U_{F}(k,\phi)  \sigma_x \neq U_F(-k,-\phi) \ .
\end{align}
However,  in the absence of a net phase, that is when $\phi_1=-\phi_2$,
this phase factor simplifies to $1$ and one gets $ \tilde{D}(-\phi_2)=\ee^{-i \phi_2/2}D(-\phi_2)$ and $ \tilde{D}(-\phi_1)=\ee^{i \phi_2/2}D(-\phi_1)$, thus
\begin{align}
	\sigma_x U_{F}^{\phi_{net}=0}(k,\phi)  \sigma_x &= B(-k)\,  S_2\,  \tilde{D}(-\phi_2)\,  B(-k)\,  S_1 \, \tilde{D}(-\phi_1)\\
	&= B(-k)\,  S_2\,  D(-\phi_2)\,  B(-k)\,  S_1 \, D(-\phi_1)\\
	&= U_{F}^{\phi_{net}=0}(-k,-\phi)  
\end{align}
so that the inversion symmetry is restored.

\subsection{Derivation of the quasienergy bands}
Let us derive here the quasienergy bands of the scattering network model with two time-steps as sketched in Fig.~\ref{fig:scatnet1}. This can be carried out analytically either by a direct diagonalization of $U_F$ or equivalently by decomposing the evolution as in Ref \cite{Wimmer2017}. Let us detail the second strategy.
Using the same terminology as in the main text, where right going arrows are denoted with $ \alpha $ and left going with $ \beta $, then the time evolution is described by 
\begin{eqnarray}\label{eq:scwim1}
	\alpha^{j+1}_{l} &=& (\cos \theta _1 \alpha^{j}_{l+1} +  i \sin \theta _1 \beta^{j}_{l+1} )\ee^{i\phi_{1} } ,\nonumber\\
	\beta^{j+1}_{l} &=& \cos \theta _1 \beta^{j}_{l-1} +  i \sin \theta _1  \alpha^{j}_{l-1} 
\end{eqnarray}
for the first step and
\begin{eqnarray}\label{eq:scwim2}
	\alpha^{j+2}_{l-1} &=& \left(\cos \theta _2 \alpha^{j+1}_{l} +  i \sin \theta _2 \beta^{j}_{l} \right)\ee^{i\phi_{2} } ,\nonumber\\
	\beta^{j+2}_{l-1} &=& \cos \theta _2 \beta^{j}_{l-2} +  i \sin \theta _2  \alpha^{j}_{l-2}
\end{eqnarray}
for the second (final) step.
Using Floquet-Bloch ansatz,
\begin{equation}\label{eq:flbl}
	\begin{pmatrix}
		\alpha^{j}_{l}\\
		\beta^{j}_{l} 
	\end{pmatrix} = \begin{pmatrix}
		A\\
		B
	\end{pmatrix} \ee^{i\varepsilon j /2}\ee^{i k l /2}
\end{equation}
and substituting Eq.\eqref{eq:scwim1} in \eqref{eq:scwim2} using Eq.\eqref{eq:flbl} gives the determinant problem 
\begin{eqnarray}\label{eq:det}
	\ee^{2i\varepsilon } - \left[\cos \theta _1\cos \theta _2\left(\ee^{i k} \ee^{i (\phi_{1}+\phi_{2})}+ \ee^{-i k}\right) -\sin \theta _1\sin \theta _2\left(\ee^{i \phi_{1}}+\ee^{i \phi_{2}}\right)\right]\ee^{i\varepsilon } +\ee^{i (\phi_{1}+\phi_{2})} = 0\ .
\end{eqnarray}
By rewriting the Eq.\eqref{eq:det}, we get the relation
\begin{eqnarray}\label{eq:disp}
	\cos\left(\varepsilon-\dfrac{\phi_{1}+\phi_{2}}{2}\right) &=& \cos\theta_{1}\cos\theta_{2}\cos\left(k+\dfrac{\phi_{1}+\phi_{2}}{2}\right) - \sin\theta_{1}\sin\theta_{2}\cos\left(\dfrac{\phi_{1}-\phi_{2}}{2}\right),\nonumber
\end{eqnarray}
that leads to 
\begin{eqnarray}
	\varepsilon_\pm(k,\phi) = \pm\cos^{-1}\left[\cos\theta_{1}\cos\theta_{2}\cos\left(-k+\dfrac{\phi_{1}+\phi_{2}}{2}\right) - \sin\theta_{1}\sin\theta_{2}\cos\left(\dfrac{\phi_{1}-\phi_{2}}{2}\right)\right]+\left(\dfrac{\phi_{1}+\phi_{2}}{2}\right)\ .
\end{eqnarray}
We can finally substitute the general form for the $ \phi $'s to $ \phi_{j} = (m_{j}/n_{j})\phi$, to get the expression
\begin{align}
	\label{eq:quasienergy}
	\varepsilon_\pm(k,\phi) =\pm\cos^{-1}\left[\cos\theta_{1}\cos\theta_{2}\cos\left(-k+\left[\dfrac{m_{1}}{n_{1}}+\dfrac{m_{2}}{n_{2}}\right]\dfrac{\phi}{2}\right) - \sin\theta_{1}\sin\theta_{2}\cos\left(\left[\dfrac{m_{1}}{n_{1}}-\dfrac{m_{2}}{n_{2}}\right]\dfrac{\phi}{2}\right)\right]+\left[\dfrac{m_{1}}{n_{1}}+\dfrac{m_{2}}{n_{2}}\right]\dfrac{\phi}{2}
\end{align}
for the quasienergy bands.

\subsection{Derivation of the group velocities}
Let us introduce the ``synthetic group velocity"  of the quasienergy band $\varepsilon_\pm$ as
\begin{eqnarray}\label{eq:slope}
	v^\pm_{\phi}(k,\phi) &\equiv&  \dfrac{\partial \varepsilon_\pm(k,\phi)}{\partial \phi}
\end{eqnarray}
Substituting the expression~\eqref{eq:quasienergy} of the quasienergy bands leads to 
\begin{eqnarray}
	v^\pm_{\phi}(k,\phi) &=& \frac{1}{2}\Delta^{+} \mp\frac{1}{2}\frac{ \Delta^{-}  \sin \theta_1 \sin \theta_2 \sin \left(\frac{1}{2} \phi  \Delta^{-} \right)-\Delta^{+}  \cos \theta_1 \cos \theta_2 \sin \left(k+\frac{1}{2} \phi \Delta^{+} \right)}{\sqrt{1-\left(\cos \theta_1 \cos \theta_2 \cos \left(k+\frac{1}{2} \phi \Delta^{+} \right)-\sin \theta_1 \sin \theta_2 \cos \left(\frac{1}{2} \phi  \Delta^{-} \right)\right)^2}}
\end{eqnarray}
where $ \Delta^{-}  \equiv \frac{m_{1}}{n_{1}}-\frac{m_{2}}{n_{2}} $ and $ \Delta^{+}  \equiv\frac{m_{1}}{n_{1}}+\frac{m_{2}}{n_{2}} $.
Similarly, the transverse group velocity is
\begin{eqnarray}
	v^\pm_{k}(k,\phi) &\equiv&  \dfrac{\partial \varepsilon_\pm(k,\phi)}{\partial k} \\
	v^\pm_{k}(k,\phi) &=& \pm\frac{\cos \theta _1 \cos \theta _2 \sin \left(k+\frac{1}{2} \phi  \Delta^{+} \right)}{\sqrt{1-\left(\cos \theta _1 \cos \theta _2 \cos \left(k+\frac{1}{2} \phi  \Delta^{+} \right)-\sin \theta _1 \sin \theta _2 \cos \left(\frac{1}{2} \phi  \Delta^{-} \right)\right){}^2}} \ .
	\label{eq:vgroup}
\end{eqnarray}

In the case $m_{1} = n_{1} = n_2 =1, m_{2} = -2$ considered in the main text, the quasienergy bands simplify  to 
\begin{equation}\label{eq:dipersion}
	\varepsilon_\pm(k,\phi) = \pm\cos^{-1}\left[\cos\theta_{1}\cos\theta_{2}\cos\left(k-\dfrac{\phi}{2}\right) - \sin\theta_{1}\sin\theta_{2}\cos\left(\dfrac{3\phi}{2}\right)\right]+\dfrac{\phi}{2}
\end{equation}
which leads to the group velocities
\begin{eqnarray}\label{eq:ftm2}
	v^\pm_{k}(k,\phi) &=& \pm\frac{\cos \theta _1 \cos \theta _2 \sin \left(k-\frac{\phi }{2}\right)}{\sqrt{1-\left(\cos \theta _1 \cos \theta _2 \cos \left(k-\frac{\phi }{2}\right)-\sin \theta _1 \sin \theta _2 \cos \left(\frac{3 \phi }{2}\right)\right){}^2}},\\
	v^\pm_{\phi}(k,\phi) &=& \frac{1}{2}\mp\frac{\frac{3}{2} \sin \theta _1 \sin \theta _2 \sin \left(\frac{3 \phi }{2}\right)+\frac{1}{2} \cos \theta _1 \cos \theta _2 \sin \left(k-\frac{\phi }{2}\right)}{\sqrt{1-\left(\cos \theta _1 \cos \theta _2 \cos \left(k-\frac{\phi }{2}\right)-\sin \theta _1 \sin \theta _2 \cos \left(\frac{3 \phi }{2}\right)\right)^2}}.
\end{eqnarray}

Note that when the quasienergy bands wind along the $\phi$ coordinate, then the quantity $ \Delta^{+}  $ is non zero. Therefore, in that case, the numerator in the expression \eqref{eq:vgroup} of the transverse group velocity 
can change  sign when varying $ \phi $, for any fixed value of $k$, thus giving rise to the wavepackets oscillations.

Along the same lines, we can calculate the motion of centre of mass from Eq.\eqref{eq:ftm2}, for arbitrary $ k $ as,
\begin{eqnarray}\label{eq:com}
	X_c(t,k) &=& \int_{0}^{\phi(t)} \dd\phi \,v_{k} (\phi(\tau),k) \left(\dfrac{\partial \phi(\tau)}{\partial \tau}\right)^{-1} ,\nonumber\\
	&=& \gamma_0\int_{0}^{t} \dd\tau \,v_{k} (\phi(\tau),k)
\end{eqnarray}
where, in the last equation, we considered that $ \phi  $ varies linearly with time with a coefficient $\gamma_0$, (see Fig. 3(a) of the main text).

\section{S2. Winding number  $\nu_\phi$ for two time-steps evolutions}

\subsection{Derivation of $\nu_\phi$.}
Let us compute the winding number $\nu_\phi$ defined in the main text, for two time-steps, where the Floquet operator $U_F(k,\phi)$ given in Eq. \eqref{eq:UF} with $\phi_1=(m_1/n_1)\phi$ and $\phi_2=(m_2/n_2) \phi$. When $ |m_1/n_1| \ne |m_2/n_2| $, then
\begin{eqnarray}\label{eq:wind1}
	\nu_{\phi} &=& \dfrac{1}{2\pi i}\int_{0}^{\Phi} \dd\phi \Tr [U_{F}^{-1}\partial_{\phi}U_{F}]\\
	&=& \dfrac{1}{2\pi i}\int_{0}^{\Phi} \dd\phi \Tr \Bigl[ D_1^{\dagger}S_1^{\dagger}B_1^{\dagger}(k) D_2^{\dagger}S_2^{\dagger}B_0^{\dagger}(k)
	\partial_{\phi}\{D_1S_1B_1(k) D_2S_2B_0(k)\}\Bigr]\\
	&=& \dfrac{1}{2\pi i}\int_{0}^{\Phi}  \dd\phi \Tr [ D_2^{\dagger}\partial_{\phi} D_2+D_1^{\dagger}\partial_{\phi}D_1]
\end{eqnarray}
where the period of $ \Phi $ of the quasienergy in $\phi$ is inferred from the analytical expression \eqref{eq:quasienergy}. More precisely, it reads
\begin{align}
	\Phi =  2\pi\ \text{LCM}\left[\frac{2}{\frac{m_1}{n_1}-\frac{m_2}{n_2}},\frac{2}{\frac{m_1}{n_1}+\frac{m_2}{n_2}}\right] 
\end{align}
where LCM stands for least common multiple.
Replacing the $D_j$'s matrices by their expression, one gets 
\begin{eqnarray}\label{eq:gen}
	\nu_{\phi} &=& \dfrac{1}{2\pi i}\int_{0}^{2\pi\ \text{LCM}\left[\frac{2}{\frac{m_1}{n_1}-\frac{m_2}{n_2}},\frac{2}{\frac{m_1}{n_1}+\frac{m_2}{n_2}}\right]}  \dd\phi \Tr [ i \dfrac{m_1}{n_1}+i \dfrac{m_2}{n_2}], \nonumber\\
	&=& 2 \hspace{0.1cm}\text{LCM}\left[\frac{1}{\frac{m_1}{n_1}-\frac{m_2}{n_2}},\frac{1}{\frac{m_1}{n_1}+\frac{m_2}{n_2}}\right]\left( \dfrac{m_1}{n_1}+ \dfrac{m_2}{n_2}\right),\nonumber\\
	&=& 2 \hspace{0.1cm}\text{LCM}\left[\frac{2 n_1 n_2}{m_1 n_2-m_2 n_1},\frac{2 n_1 n_2}{m_1 n_2+m_2 n_1}\right]\left( \dfrac{m_1}{n_1}+ \dfrac{m_2}{n_2}\right),\nonumber\\
	\nu_\phi&=& \dfrac{\Phi}{2\pi}\left( \dfrac{m_1}{n_1}+ \dfrac{m_2}{n_2}\right).
\end{eqnarray}

\subsection{Relation between $\nu_\phi$ and the stationary points of the Bloch oscillations}
Consider a situation where the quasienergy bands wind along $ \phi $, and let us count the number of stationary points $\frac{\dd X_c}{\dd t}$ in Eq.\eqref{eq:com} over one Bloch period of oscillation. These points are determined by the vanishing of the group velocity $v_k$. Therefore, it suffices to find the number of roots in $\phi$ of Eq.\eqref{eq:vgroup}, which are given by
\begin{eqnarray}\label{eq:stationary}
	\cos \theta _1 \cos \theta _2 \sin \left(\frac{1}{2} \phi  \Delta^{+} \right) = 0
\end{eqnarray}
This leads to 
\begin{eqnarray}
	\frac{1}{2} \phi  \Delta^{+} &=& p\pi, \qquad p \in\mathds{Z} \nonumber\\
	\phi  &=& 2p\pi \dfrac{1}{\Delta^{+}}.
\end{eqnarray}
which can be expressed in terms of the winding number  given by Eq.\eqref{eq:gen} as
\begin{eqnarray}\label{eq:stationary_nu}
	\phi  &= p \dfrac{\Phi}{\nu_{\phi}}.
\end{eqnarray}
Hence, over one oscillation period, $ p $ takes values from the set $ \{ 1, ...,|\nu_{\phi}|\} $, and thus the group velocity vanishes $ \nu_{\phi} $ times. Then, following the same lines, one can easily check that the second derivative also vanishes at these same points. There are therefore $\nu_\phi$ turning points per period of Bloch oscillation.

\begin{figure}[!htb]
	\includegraphics[width=0.7\linewidth]{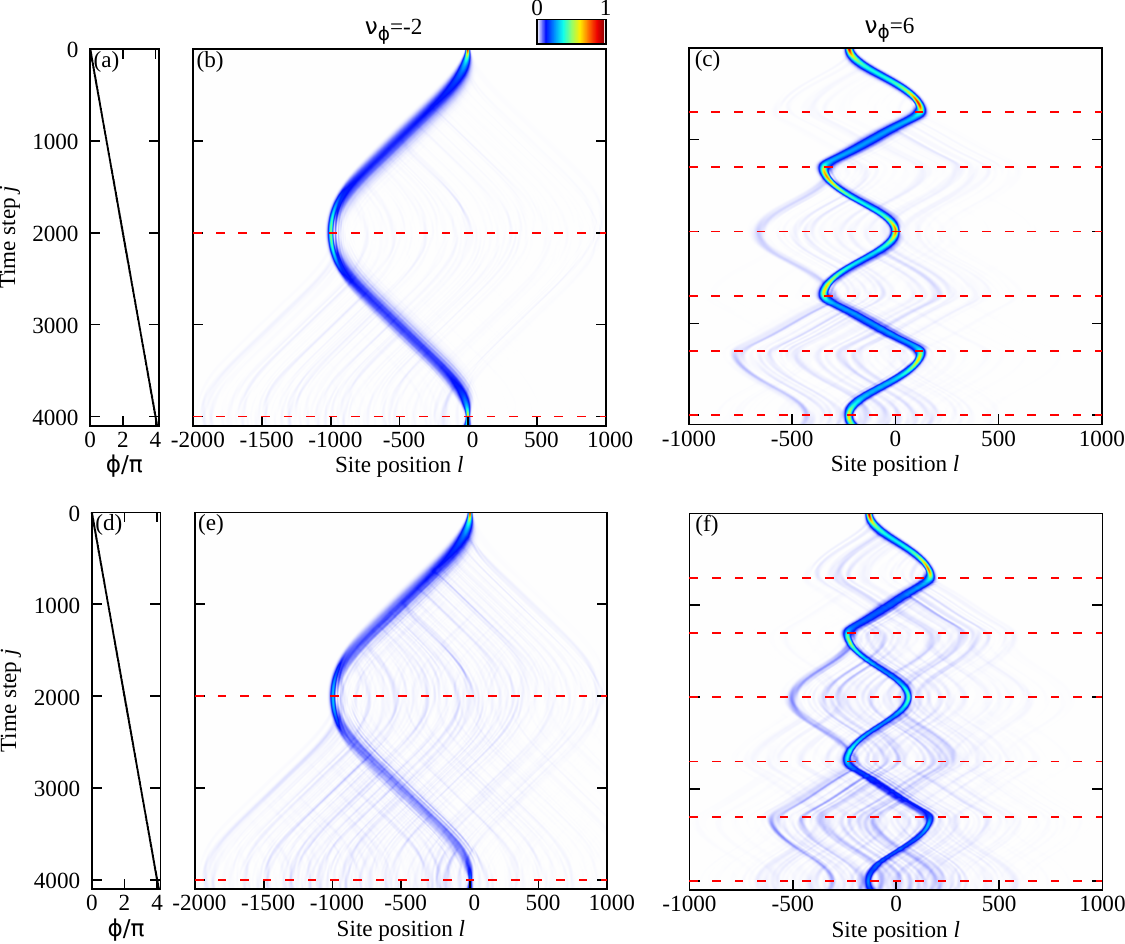}
	\caption{Intensity ($|\alpha_l^j|^2 + | \beta_l^j|^2$) of a wavepacket, injected with a Gaussian shape when the values of $\theta_1$ and $\theta_2$ have a random disorder that is periodic with the Floquet period ($\theta_{i}+\delta_{i}$ with $i=1,2$ and $\delta_{i}$ a uniform distribution between $[-A,+A]$). Top: $A=0.02$, bottom: $A=0.04$. Left column: $\nu_{\phi}=-2$. Right column: $\nu_{\phi}=6$. Other parameters have the same value than in the Fig.~3 of the main text. For visibility, horizontal red dashed lines are placed at each turning point. }
	\label{fig:noise}
\end{figure}

\subsection{Robustness of $\nu_\phi$ under disorder}
The Bloch oscillations emerged due to the underlying discrete translational symmetry of the lattice. However, these Bloch oscillations have topological origin where different values of $\nu_\phi$ constitute the family. To check the robustness of these $\nu_\phi$, we introduce a disorder in the coupling parameters ($ \theta_{i}$ with $i=1,2$); thus we break the discrete translational symmetry of the lattice. This is achieved by putting a disorder as $ \theta_{i}+\delta_{i} $, where $ \delta_{i} $ is a random number with a uniform distribution between the interval  $[-A, +A]$.
The dynamics of the wavepacket in these conditions is shown in figure~\ref{fig:noise}, for the case of $\nu_{\phi}=-2$ and $\nu_{\phi}=6$, and for a noise amplitude  $A$ of 0.02 and 0.04 (the other parameters are same as in Fig.~3 of the main text).\\
Besides the main oscillation, it appears, some ``sub-branches" that seem to leave the main wavepacket.  This effect is related to the presence of two bands in the system. The initial condition has been chosen to excite only one band of the system, however, with the variation of $\theta_i$ (induced by the noise term), the solution is not anymore restricted to two bands but becomes a multi-band problem, as a result, the other bands are also excited. \\

It induces the creation and evolution of ``secondary" wavepackets. However, even with these wavepackets, the topological invariant $ \nu_{\phi} $ is preserved, since we can see that in all these wavepackets, the turning points appear at the same time ($ y $-axis), and thus after one Bloch oscillation period, they have the same number of turning points, i.e. the same $\nu_{\phi}$.

\section{S3. Fictitious electric field in the network model}\label{sec:s4}
\subsection{Gauge transformation from a uniform electric field to winding bands with an adiabatic increase of $\phi$}
As pointed out by M. Wimmer et al. in Ref.~\cite{Wimmer2015} in the case of a single step model with an adiabatic increase of the phase factor $\phi(j)= \gamma_0\, j$, the gauge transformation:
\begin{equation}
	\begin{aligned}\label{eq:gaugetransf}
		\alpha_{l}^{j+1}&= \tilde{\alpha}_{l}^{j+1} \ee^{-\frac{ilj\gamma_0}{2}+\frac{ij^2\gamma_0}{4}-\frac{ij\gamma_0}{4}}\\
		\beta_{l}^{j+1}&= \tilde{\beta}_{l}^{j+1} \ee^{-\frac{ilj\gamma_0}{2}+\frac{ij^2\gamma_0}{4}-\frac{ij\gamma_0}{4}}
	\end{aligned}   
\end{equation} 
results in a set of equations in which the phase factor does not depend anymore on the time step, but presents a uniform gradient of phase~:
\begin{equation}
	\begin{aligned}\label{eq:Bloch}
		\tilde{\alpha}_{l}^{j+1}&=(\cos \theta_{j} \tilde{\alpha}_{l+1}^{j} + i \sin \theta_{j} \tilde{\beta}_{l+1}^{j}) \ee^{\frac{i\gamma_0 l}{2}}\\
		\tilde{\beta}_{l}^{j+1}&=(i \sin \theta_{j} \tilde{\alpha}_{l-1}^{j} + \cos \theta_{j} \tilde{\beta}_{l-1}^{j}) \ee^{\frac{i\gamma_0 l}{2}} \ .
	\end{aligned}   
\end{equation}

This set of equations corresponds to a scattering network subject to a homogeneous spatial phase gradient $V= E\cdot l$, where $E=\gamma_0/2$ can be interpreted as a homogeneous electric field along the $l$ direction. When considering an initial wavepacket, the time evolution results in standard Bloch oscillations with period $T=2\pi/E=4\pi/\gamma_0$.

The same gauge transformation can be applied to each of the two-steps of the model with $n=2$ discussed in the main text subject to an adiabatic increase of $\phi(j)=+\gamma_0 j$ (Fig.~3 of the main text).

Recall that in the first step $\phi_1(j)=(m_1/n_1)\gamma_0 j$ and in the second step $\phi_2(j)=(m_2/n_2)\gamma_0 j$. The transformation (\ref{eq:gaugetransf}) results in a set of the Bloch-like Eqs.\eqref{eq:Bloch} for each of the two Floquet steps, each set characterized by a constant electric field in space. In the first step, the electric field is $E_1=(m_1/n_1)\gamma_0/2$, and in the second step is $E_2=(m_2/n_2)\gamma_0/2$. 
Therefore, we get back the Bloch oscillation picture in this case with an electric field that alternates between $E_1$ and $E_2$ at each subsequent step. The period $T_B$ of the oscillations  can be computed from the average electric field $(E_1+E_2)/2$ over a full Floquet cycle.

The above discussion can also be simply understood from basic classical electrodynamics arguments\cite{Krieger1986,Zak1988}. Indeed, in its most general form, an electric field can be expressed as $\mathbf{E}=-\mathbf{\nabla} V +\partial \mathbf{A}/\partial t$.
The Eq.\eqref{eq:gaugetransf} is the gauge transformation that transforms a gradient of spatial potential $V$, to a time-varying vector potential $A$.

\subsection{Fictitious uniform electric field from a fictitious  vector potential}

The above discussion can also seen from the Floquet operator Eq.\eqref{eq:UF} using the simplification as in Eq.\eqref{eq:UF_net},
\begin{eqnarray}\label{eq:floquet_wim}
	U_{F}(k,\phi) 
	= \ee^{i(\phi_{1}+\phi_{2})/2} \, B(k) \, S_{2} \, \tilde{D}(\phi_{2})\, B(k) \, S_{1}\,  \tilde{D}(\phi_{1}) \  ,
\end{eqnarray}
where $ B $ and $ \tilde{D} $ are defined in Eq.\eqref{eq:bk} and Eq.\eqref{eq:dphi}.
Then $ U_{F} $ can be further simplified by combining two diagonal matrices $ \tilde{D} $ and $ B $ together, (also used in eq.(3) in the main text) 
		\begin{align} \label{eq:floquet_def}
		U_{F}(k,\phi) &= \ee^{i(\phi_{1}+\phi_{2})/2}\, T_2\,S_{2}\, T_1\, S_{1} \\
				T_j& =\begin{pmatrix}
			\ee^{i  \tilde{k}_{j}} & 0\\
			0 & \ee^{-i  \tilde{k}_{j}}
		\end{pmatrix}\nonumber
\end{align}
where $ \tilde{k}_{j} = k-\phi_{j} $.
This form of the Floquet operator can be thought as describing a 1D lattice that is periodically driven in the presence of a time-varying vector potential $ A(t) = E t $. The period $ T $ of this driving consists of two-steps where the vector potential redefines the Bloch momentum via Peierls' substitution. In the first step, for some fictitious charge $ q $, $ \phi_{1} = q A_{1} = q E_{1} t$ that generates a fictitious electric field of magnitude $ E_{1} $. Similarly, during the second step, $ \phi_{2} = q A_{2} = q E_{2} t$. This electric field translates in our case as $ \phi_{1} = -2\phi = -2q E$ and $ \phi_{2} = +\phi = +q E$. Thus, it gives rise to a net electric field $ E_{1}+E_{2}\ne 0 $, which is responsible for the Bloch oscillations.

\section{S4. Extended network model for quasienergy winding in $k$}
\subsection{Time-step evolution equations for the scattering model with quasienergy winding in $k$}

\begin{figure}[t]
	\centering
	\includegraphics[width=0.75\columnwidth]{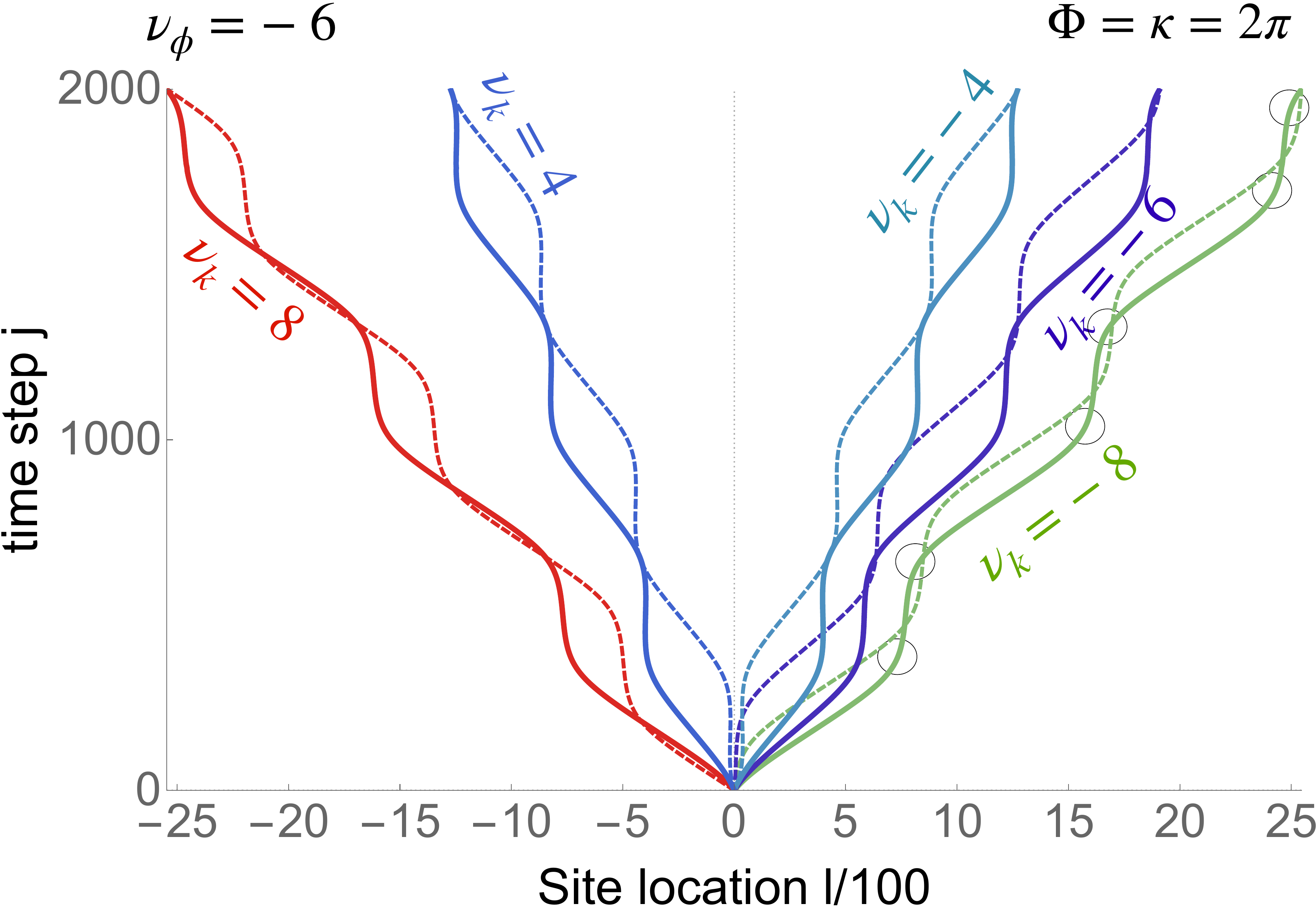}
	\caption{Trajectories in a two-steps model for bands with $\nu_{\phi}=-6$ for $ \theta_{1} = \pi/4, $ $ \theta_{2} = \pi $, and different values of $\nu_k$, ranging from $\nu_k = +8$ to $\nu_k = -8$, where the thick curves are for $ k = 0 $, and the dashed curves are for $ k = 1 $. Thus, changing $ k $ merely shifts the curve in vertical direction. Moreover, $|\nu_{\phi}| = 6$ can still be read even in this case of winding in both $ k $ and $ \phi $ from the number of turning points, irrespective of the initial value of $ k $, as shown with little circles for $ \nu_k = -8 $ (in green). Here, $\phi$ increases as $\phi(j)=\gamma_0 j$, with $\gamma_0=2\pi/2000$}
	\label{fig:drift_k_phi}
\end{figure}

In the final part of the main text we introduce a model with long range hoppings that results in the winding of the quasienergy bands in the $k$ direction. The model is sketched in Fig.~4(a) of the main text. The corresponding time step evolution equations are given by~:
\begin{eqnarray}
	\label{eq:timestepgeneral1}
	\alpha_{l}^{j+1}&=(\cos \theta_{j} \alpha_{l+l_{2}}^{j} + i \sin \theta_{j} \beta_{l+l_{2}}^{j}) \ee^{i \phi_{j}}\nonumber\\
	\beta_{l}^{j+1}&=(i \sin \theta_{j} \alpha_{l+l_{0}}^{j} + \cos \theta_{j} \beta_{l+l_{0}}^{j}) \, ,\\\nonumber\\
	\label{eq:timestepgeneral2}
	\alpha_{l+l_{3}}^{j+2}&=(\cos \theta_{j+1} \alpha_{l}^{j+1} + i \sin \theta_{j+1} \beta_{l}^{j+1}) \ee^{i \phi_{j+1}}\nonumber\\
	\beta_{l+l_{3}}^{j+2}&=(i \sin \theta_{j+1} \alpha_{l+l_{4}}^{j+1} + \cos \theta_{j+1} \beta_{l+l_{4}}^{j+1}) \, .
\end{eqnarray}
Here $ l_{j} $ is the link connecting the scattering nodes at time step $ j+p-1 $ to $ j+p $, for some integer $ p $. These $ l_j $'s  can be defined in terms of $ r_j/s_j $ as
\begin{eqnarray}\label{eq:relation}
	\dfrac{r_1}{s_1} = \dfrac{l_{2}-l_{3}}{2},\qquad
	\dfrac{r_2}{s_2} = \dfrac{l_{2}-l_{3} + l_{0}}{2} \ ,
\end{eqnarray}
If $\phi_1=-\phi_2$, then there is no winding in $\phi$, and the bands only wind in $k$. By combining both windings in $k$ and $\phi$, the drift and pseudo-oscillations of a wavepacket can be engineered together. Trajectories for the center of mass are shown in Fig.~\ref{fig:drift_k_phi} for a fixed value of $ \nu_{\phi} = -6 $, and different values of $ \nu_{k} $, where thick curves are evaluated for a wavepacket centered at $k = 0 $, and dashed curves for $ k = 1$. The value of $ \nu_{\phi} $ can still be inferred from the number of turning points, as shown for $ \nu_{k} = 8 $.

\subsection{Derivation of the winding number $\nu_k$}
For an arbitrary winding number in $ k $ and $ \phi $, the Floquet operator reads
\begin{eqnarray}\label{eq:s2g}
	U_{F}(k,\phi) &=& B_{0}(k_{2})\,  S_{2} \, D(\phi_2) \, B_{1}(k_{1})\,  S_{1} \, D(\phi_1)
\end{eqnarray}
where, $ k_{j} \equiv \left(\dfrac{r_j}{s_j}\right) k$ and  $ \phi_{j} \equiv \left(\dfrac{m_j}{n_j}\right) \phi$.
This gives the quasienergies $ \varepsilon_{\pm}(k,\phi) $
\begin{equation}\label{eq:winkp}
	\varepsilon_{\pm}(k,\phi) = \pm\cos^{-1}\left[\cos\theta_1 \cos\theta_2 \cos \left(\delta^{-}\frac{k}{2}+ \Delta^{+}\frac{\phi}{2}\right)-\sin\theta_1 \sin\theta_2 \cos \left(\delta^{+}\frac{k}{2}- \Delta^{-}\frac{\phi}{2}\right)\right]
	+\delta^{+}\frac{k}{2}+\Delta^{+}\frac{\phi}{2}\, ,
\end{equation}
where $ \delta^{\pm} \equiv \frac{r_1}{s_1}\pm\frac{r_2}{s_2} $.
In the absence of a quasienergy winding along  $ \phi $ (or $ k $), the corresponding $ \Delta^{+}\ (\delta^{+})  $ terms vanish.
The group velocity can then be derived exactly as
\begin{equation}
	v_{g \pm}(\phi,k) = \dfrac{1}{2}\delta^{+}\pm\frac{1}{2}\frac{\delta^{+}\sin\theta_1\sin\theta_2\sin \left(\delta^{+}\frac{k}{2}-\Delta^{-}\frac{\phi}{2}\right)-\delta^{-}\cos\theta_1\cos\theta_2\sin \left(\delta^{-}\frac{k}{2}+ \Delta^{+}\frac{\phi}{2}\right)}{\sqrt{1-\left(\cos\theta_1\cos\theta_2\cos \left(\delta^{-}\frac{k}{2}+ \Delta^{+}\frac{\phi}{2}\right)-\sin\theta_1\sin\theta_2\cos \left(\delta^{+}\frac{k}{2}- \Delta^{-}\frac{\phi}{2}\right)\right)^2}}.
\end{equation}

Likewise, the winding number in $ k $ can be computed similar to $\nu_\phi$ as
\begin{eqnarray}\label{eq:gen1}
	\nu_{k} &\equiv& \dfrac{1}{2\pi i} \int_{0}^{\kappa}  \dd k \Tr [U_F^{-1} \partial_k U_F ] \\
	&=& \dfrac{1}{2\pi i}\int_{0}^{\kappa}  \dd k \Tr [ i \frac{r_1}{s_1}+i \frac{r_2}{s_2}], \nonumber\\
	&=& 2 \hspace{0.1cm}\text{LCM}\left[\frac{1}{\frac{r_1}{s_1}-\frac{r_2}{s_2}},\frac{1}{\frac{r_1}{s_1}+\frac{r_2}{s_2}}\right]\left( \frac{r_1}{s_1}+ \frac{r_2}{s_2}\right),\nonumber\\
	&=& 2 \hspace{0.1cm}\text{LCM}\left[\frac{2 s_1 s_2}{r_1 s_2-r_2 s_1},\frac{2 s_1 s_2}{r_1 s_2+r_2 s_1}\right]\left( \frac{r_1}{s_1}+ \frac{r_2}{s_2}\right),\nonumber\\
	\nu_k&=& \dfrac{\kappa}{2\pi}\left( \frac{r_1}{s_1}+ \frac{r_2}{s_2}\right)
\end{eqnarray}
where $ \kappa = 2\pi\ \text{LCM}\left[\frac{2}{\frac{r_1}{s_1}-\frac{r_2}{s_2}},\frac{2}{\frac{r_1}{s_1}+\frac{r_2}{s_2}}\right]$.

\section{S5. Relation between the winding number $ \nu_{k} $ and the quantized drift $\Delta x$}

Let us introduce the mean current over a Floquet period $T$ as
\begin{align}
	\label{eq:mean}
	J\equiv \int_0^T \frac{\dd t}{T} j(t)
\end{align}
that we express in terms of the instantaneous current $ j(t)$ 
\begin{align}
	\label{eq:current_k}
	j(t) = \int_{0}^{\kappa}  \frac{\dd k}{\kappa} \langle\psi(k,t)|\dfrac{\dd x(t)}{\dd t}|\psi(k,t)\rangle,
\end{align}
where $|\psi(k,t)\rangle$ is an arbitrary evolving Bloch state, i.e. $|\psi(k,t)\rangle=U(k;t,0)|\psi(k,0)\rangle$ with $U(k;t,0)$ the Block evolution operator from time $t=0$ to arbitrary time $t<T$. Rewriting $U(k;t,0)=U(k;t,T)U(k;T,t)$, and using the relation $i \frac{\partial U_F}{\partial k} = \int_0^T \dd t\, U(k;T,t)\,\frac{\partial H(k,t)}{\partial k} \,U(k;t,0)$, where $ H(k,t+T) = H(k,t) $ is the periodically driven Bloch Hamiltonian, the mean current can be written in terms of the Floquet operator only
\begin{align}
	\label{eq:mean2}
	J = -\frac{2\pi /\kappa}{T} \int_0^{\kappa} \frac{\dd k}{2\pi i} \bra{\psi(k,0)}U_F^{-1}\frac{\partial U_F}{\partial k}  \ket{\psi(k,0)}
\end{align}
Equivalently, one assigns a mean displacement $\Delta x =T J$ to this current.

\subsection{Adiabatic regime}
Consider an instantaneous eigenstate $\varphi^{(n)}(k,t)$ of $ H(k,t)$, such that $\psi(k,0)= \varphi^{(n)}(k,t=0)$. In the adiabatic limit, $ \varphi^{(n)}(k,t) $ remains an eigenstate of $ H(k,t)$ at each time. After a cycle $t:0\rightarrow T$, $ \varphi^{(n)}(k,0) $ can only acquire a phase, which is by definition the quasienergy $\epsilon_n T$. It is thus an eigenstate of the Floquet operator, which therefore allows the spectral decomposition $U_F= \sum_n^N \exp(-i \epsilon_n T)|\varphi^{(n)}(k,0)\rangle \langle \varphi^{(n)}(k,0) |$, so that the mean current \eqref{eq:mean2} simply becomes
\begin{align}
	J^{(n)}_{ad} = -\frac{2\pi/\kappa}{T} \int_0^\kappa \frac{\dd k}{2\pi} \frac{\partial \varepsilon_n}{\partial k}
\end{align}
where the dimensionless quasienergy $\varepsilon_n = -\epsilon_n\, T$ corresponds to that of the main text. The adiabatic pumped current is quantized in terms of the quasienergy winding numbers along the $k$ direction, as found in Ref~\cite{Kitagawa2010}. As pioneered by Thouless \cite{Thouless1983}, this quantization can be consistently rephrased in terms of the Chern numbers $C_n$ of the adiabatically driven eigenstates $\varphi^{(n)}(k,t)$ that defined a $U(1)-$fiber bundle over the two-dimensional torus span by $(k,t)$, assuming the instantaneous energy band $E_n(k,t)$ (the eigenvalue of $H(k,t)$) remains well separated from the other bands. One way to see the connection between the two topological points of view consists in identifying the quasienergy in terms of the dynamical phase and the geometrical Berry phase
\begin{align}
	\epsilon_n T = E_n T + i \int_0^T \dd t \langle \varphi^{(n)}(k,t) | \partial_t | \varphi^{(n)}(k,t) \rangle \ .
\end{align}
Taking, the ``winding''  of this expression, that is applying $\int \dd k \partial_k$ yields
\begin{align}
	\label{eq:wind}
	-\int \frac{\dd k}{2\pi} \partial_k \varepsilon_n =  i \int \frac{\dd k}{2\pi} \int_0^T\partial_k \dd t \langle \varphi^{(n)}(k,t) | \partial_t | \varphi^{(n)}(k,t) \rangle
\end{align}
since the instantaneous energy band $E_n(k,t)$ cannot wind along $k$. Inserting the relation $\partial_k \langle \varphi| \partial_t \varphi \rangle = \langle \partial_k \varphi | \partial_t \varphi\rangle - \langle \partial_t \varphi| \partial_k \varphi \rangle + \partial_t \langle \varphi | \partial_k \varphi \rangle$, into the  right-hand side of \eqref{eq:wind}, the quasienergy winding reads
\begin{align}\label{eq:drift_adiabatic_2}
	-\int \frac{\dd k}{2\pi} \partial_k \varepsilon_n  = \frac{1}{2\pi} \int_0^T\int \dd k F_{k,t}^{(n)} + \int_0^T \partial_t Z^{(n)}(t)
\end{align}
where $F_{k,t}^{(n)}$ is the Berry curvature and $Z^{(n)}(t)$ is the  time-dependent Zak phase of the instantaneous state $\varphi^{(n)}$, i.e. $Z^{(n)}(t)\equiv i \int \dd k \langle \varphi^{(n)}(k,t) | \partial_k \varphi^{(n)}(k,t)\rangle$. After an adiabatic cycle, one has $Z^{(n)}(T)=Z^{(n)}(0)$, which leads to the relation between the winding number $\nu_k^{(n)}$ of the quasienergy band $n$ in the $k$ direction and the Chern number $C_n$ of the adiabatically periodically driven Bloch eigenstate $\varphi^{(n)}(k,t)$
\begin{align}
	-\int \frac{\dd k}{2\pi} \partial_k \varepsilon_n =C_n \ .
\end{align}
When the $\alpha$ lowest bands are filled, the adiabatic pumped current reads
\begin{align}
	\bar{J}_\alpha = \sum_{n=1}^{\alpha} J_{ad}^{(n)} = \frac{2\pi/\kappa}{T}\sum_{n=1}^{\alpha} C_n
\end{align}
in agreement with the famous Thouless result on adiabatic pumping. Clearly, if all the bands are filled, then $\bar{J}_N=\nu_k=0$ owing to the vanishing sum of the Chern numbers over all the bands.

\subsection{Non-adiabatic regime}
We now consider the case where the instantaneous eigenstates $ \varphi^{(n)}(k,t) $ do not remain eigenstates during the evolution, so that the total mean current $\bar{J}_N$ reads
\begin{align}
	\bar{J}_N &= \sum_{n=1}^{N} \int_0^T \frac{\dd t}{T}  \int_0^{\kappa} \frac{\dd k}{\kappa}  \langle\varphi^{(n)}(k,t)|\dfrac{\dd x(t)}{\dd t}|\varphi^{(n)}(k,t)\rangle \\
	&= -\frac{2\pi /\kappa}{T} \int_0^{\kappa} \frac{\dd k}{2\pi i} \tr[U_F^{-1}\frac{\partial U_F}{\partial k} ]\\
	& = -\frac{2\pi /\kappa}{T} \nu_k
\end{align}
with $\nu_k \in \mathbb{Z}$ is the winding number of the map $k\in S^1\rightarrow U_F \in U(N)$, and whose another expression is given by Eq.(9) of the main text.
Moreover, since $\tr U_F^{-1} \frac{\partial U_F}{\partial k}  = \frac{\partial }{\partial k} \ln \det U_F$, this winding number reads 
\begin{align}
	\nu_k =  \sum_{n=1}^{N} \int_0^{\kappa} \frac{\dd k}{2\pi} \frac{\partial (-\epsilon_n T)}{\partial k}=\sum_{n=1}^{N} \int_0^{\kappa} \frac{\dd k}{2\pi} \frac{\partial \varepsilon_n}{\partial k}
\end{align}     
so that the mean displacement $ \Delta x =T \bar{J}_N$, after $P$ periods $T$, can be expressed in terms of the sum of the winding of the quasienergies of all the bands
\begin{align}
	\Delta x = -P \frac{2\pi}{ \kappa} \sum_{n=1}^{N} \int_0^{\kappa} \frac{\dd k}{2\pi} \frac{\partial \varepsilon_n}{\partial k}
\end{align}

\bibliography{Windingbib}
\end{document}